\documentclass[11pt]{article} 

\usepackage{fullpage}
\usepackage{graphicx}
\usepackage{upref}
\usepackage{enumerate}
\usepackage{latexsym}
\usepackage{pdfpages}
\usepackage[bookmarks]{hyperref}
\usepackage{color,graphics}
\usepackage{graphicx}
\usepackage{comment} 
\usepackage{caption}
\usepackage{subcaption}
\usepackage{wrapfig}
\usepackage{braket}

\usepackage{amssymb}
\usepackage{amsmath}
\usepackage{bbm}
\usepackage{amsthm}
\usepackage{mathrsfs}
\usepackage{mathtools}

\usepackage{epsfig,graphicx}
\usepackage{algorithmic}
\usepackage[ruled,linesnumbered]{algorithm2e}
\usepackage{amsfonts}
\usepackage{tikz} 
\usepackage{varioref}
\usepackage[ansinew]{inputenc}
\usepackage{qcircuit}
\usepackage{authblk}
\usepackage{multirow}
\usepackage{lineno}

\theoremstyle{plain}
\newtheorem{theorem}{Theorem}
\theoremstyle{plain}
\newtheorem{lemma}[theorem]{Lemma}
\theoremstyle{plain}

\newtheorem{corollary}[theorem]{Corollary}
\theoremstyle{plain}

\theoremstyle{plain}

\theoremstyle{plain}

\theoremstyle{definition}

\newtheorem{definition}[theorem]{Definition}
\theoremstyle{definition}

\theoremstyle{remark}

\theoremstyle{definition}

\AtBeginDocument{}

\newcommand{\fld}{\mathbb{F}}

\newcommand{\conj}[1]{\overline{#1}}
\newcommand{\indx}{\mathcal{I}}

\begin{document}

 \title{ Graphical and algebraic methods for Boolean factoring }

\author[1]{Simon Martiel \thanks{martiel@ibm.com}}
\author[2]{Priyanka Mukhopadhyay \thanks{mukhopadhyay.priyanka@gmail.com, Priyanka.Mukhopadhyay@ibm.com}}

 \affil[1]{IBM Quantum, IBM France Lab., Saclay, France}
 \affil[2]{IBM Quantum, IBM T.J.Watson Research Center, Yorktown Heights, NY 10598}

\date{}
 \maketitle
 
 \begin{abstract}

The problem of factoring Boolean polynomials has significant applications in both classical and quantum computing technology. In this paper we have developed novel algorithms for factoring both ESOP and SOP expressions. Our aim is to optimize the AND-count. The AND-count plays a key role in determining the number of AND and Toffoli gates required to implement a reversible function with classical and quantum circuits, respectively. The first type of algorithms that we develop, are graphical. We reduce the problem of Boolean factoring to covering a biparitite graph with bicliques, and so optimizing the number of bicliques required to cover the bipartite graph, leads to reduced number of factors, and hence AND-count. The second type of algorithm is algebraic, and is derived from multivariate Horner method. We have compared the performances of our algorithms with existing popular methods like EXORCISM-4 and EPOEM2, on random functions of up to 12 variables. We have observed that our multivariate Horner method is substantially faster, while our biclique-based method achieves the maximum AND-count reduction. In fact, compared to EXORCISM-4 our biclique based method achieves up to 5 times reduction in AND-count.

 \end{abstract}

\section{Introduction}
\label{sec:intro}

The problem of polynomial factorization has been widely studied in algebra and it has many applications \cite{2003_GG}. Specifically, the problem of factoring Boolean polynomials has received much attention because of its significant application in areas like classical and quantum computer technology. Boolean polynomials are multilinear polynomials over the finite field of order 2. There are four main Boolean operators - negation, AND ($\wedge$ or $.$), OR ($\vee$) and XOR ($\oplus$). Most often, Boolean polynomials or expressions are expressed as sum of monomials, that are product (AND) of Boolean variables. The polynomial is said to be in SOP (Sum Of Product) or ESOP (XOR Sum Of Product) form, if the sum is OR or XOR, respectively. 

Both SOP and ESOP forms have been widely used to represent and manipulate Boolean functions. They have their own merits and demerits. Classically, the XOR gate is more costly to implement than the NAND and NOR gates. For efficient implementation we need to minimize the Boolean expressions, and the exact minimization of arbitrary ESOPs is practical only for functions with up to 5 input variables, and special classes of functions with up to 10 variables \cite{1993_S}. But good minimzation algorithms for SOP expressions can be found in \cite{1955_Q, 1974_HCO, 1984_BHMS, 1984_S}. However, for many Boolean functions, the number of cubes (summands) in minimal ESOPs is less than the number of cubes in minimal SOPs \cite{1996_SF}. Also, many classical logic can be implemented in practice with less area and delay, with the selective use of XOR gates \cite{2000_YCS, 2001_MSP, 2017_SP}. Programmable logic arrays (PLAs) using XOR gates require less hardware \cite{1993_S2}. For fully homomorphic encryption (FHE) the XOR gate is considered much cheaper than AND gate, and do not increase the noise during the computation (\cite{2019_TSAM}).Further, the XOR gate has excellent testability properties leading to efficient methods for automatic test pattern generation and design for test \cite{2002_KHP}. In quantum circuit synthesis flows, ESOP expressions are used as intermediate representations. The complexity of these expressions are determined by the number of gates in the resulting quantum circuits. The reversible functions implemented by Clifford+Toffoli quantum circuits can be expressed as ESOP expressions \cite{2011_DFW, 2017_SRWM, 2018_MSRetal, 2019_MSCetal, 2019_MSEetal}. 

Factoring is one of the methods by which we can re-write a Boolean expression in a way that it possibly leads to its more efficient implementation. Factoring a polynomial is the procedure of deriving an equivalent paranthesized algebraic expression or factored form. For example, the ESOP polynomial $f = ab\oplus bc\oplus ca$, can be factored as
\begin{eqnarray}
    f = ab\oplus bc\oplus ca = b(a\oplus c)\oplus ca.   \label{eqn:factor1} 
\end{eqnarray}
In the above expressions, for clarity, we have omitted the AND symbol for product of variables or expressions. A single polynomial can have more than one equivalent factored forms. For example, another factored form of the above polynomial is
\begin{eqnarray}
    f = (a\oplus c)(b\oplus c)\oplus c. \label{eqn:factor2}
\end{eqnarray}
Depending on applications, an optimum factored form can refer to one with the shortest length or minimum number of AND terms. Generating an optimum factored form is an NP-hard problem \cite{2005_MG}. The minimum number of AND or product operators required to express a Boolean expression in SOP or ESOP form, is referred to as its \emph{multiplicative complexity}. For example, Equation \ref{eqn:factor2} with multiplicative complexity 1, shows an optimum factored form of the ESOP polynomial $f$. Most of the algorithms for Boolean factoring are heuristic and they are either based on algebraic division \cite{1987_B, 1982_B, 2004_BRSW, 2007_KPVetal}, or Boolean division \cite{1988_K, 1992_HS, 1999_CC, 2002_SS}, or graph partitioning \cite{2005_MG, 2006_GMR}.

 In classical logic synthesis, the multiplicative complexity of a function is equal to the minimum number of AND gates required to implement it. Computing the multiplicative complexity of arbitrary Boolean function is intractable \cite{2014_F}. For some classes of Boolean functions the exact multiplicative complexity is known, for example, all Boolean functions with up to 6 inputs \cite{2019_CSP} and all symmetric Boolean functions \cite{2008_BP}. In the implementation of many cryptographic protocols it is crucial to minimize the number of AND gates, for example, zero-knowledge protocols, fully homomorphic encryption (FHE) and secure multi-party computation (MPC)\cite{2015_ARSetal, 2016_GMO, 2017_CDGetal, 2024_YM, 1986_Y, 2015_SHSetal, 2008_KS, 2019_TSAM}. Lower multiplicative complexity of a function corresponds to higher vulnerability to algebraic attacks \cite{2014_TSP, 2011_CHM}. The cost of general-purpose protection against side-channel attacks, grows with the number of AND gates \cite{2011_CHM}. Motivated by all these applications, many heuristic algorithms have been developed to optimize the multiplicative complexity of Boolean circuits \cite{2008_BP, 2019_TSAM, 2013_BMP, 2019_CSP, 2020_S, 2020_TSRetal}.

In quantum computing, the multiplicative complexity of a reversible function determines the minimum number of Toffoli gates required to implement it with a Clifford+Toffoli quantum circuit \cite{2017_PRS, 2019_MSCetal}. This is especially useful for the construction of quantum oracle circuits \cite{2019_MSCetal, 2019_MSEetal, 2020_HS, 2023_SBBetal, 2025_Mqram}, that play a significant role in many quantum algorithms \cite{1996_G, 2019_BDLK, 2020_AGGW, 2021_OCKK}. And hence it is essential to efficiently implement them in order to recover the claimed quantum speed-ups \cite{2021_BLHetal}. In the compilation of Trotter circuits for Hamiltonian simulation, Toffoli gates can be used to trade-off more rotation gates, that are more costly to implement fault-tolerantly \cite{2023_MWZ}. Toffoli gates have also been used to implement some special signature matrices, for example in \cite{2026_HMAetal}. Both the last two applications include the implementation of some reversible functions, and hence a good factoring algorithm can potentially lead to reduced Toffoli-count. 

Here we briefly mention that any unitary can be implemented with arbitrary accuracy by a discrete, finite, universal gate set \cite{1997_K, 2005_DN}. To protect the computational results from undesired and irrecoverable errors, it is desired that a universal gate set has a fault-tolerant implementation in any quantum error correction code. The Clifford+T is the most popular fault-tolerant universal gate set, closely followed by Clifford+Toffoli. Most researchers prefer to implement their algorithms, at least in theory, with these two gate sets. The Toffoli gate can be implemented with the Clifford+T gate set, and hence reducing the Toffoli count/depth of a quantum circuit implies reducing the T-gate count/depth, as well. But the converse is not true \cite{2024_Mtof}. In most quantum error correction schemes, the cost of implementing a non-Clifford gate like the T or Toffoli, is much more than the cost of implementing a Clifford gate \cite{2005_BK, 2020_BCHK}. Hence for efficient implementation of reversible functions on a quantum computer, it is important to reduce its multiplicative complexity. 

\subsection{Our contributions}

In this paper we have developed novel algorithms for factoring both ESOP and SOP expressions, with the aim of optimizing the AND-count of the factored expressions. The first type of algorithm is graphical. We show that factoring of Boolean expressions can be reduced to the problem of covering a bipartite graph with bicliques. So optimizing the number of bicliques required to cover a bipartite graph, will lead to reduced number of factors, and hence AND-count. The second algorithm is derived from multivariate Horner method. We have compared the performance of our algorithms on random Boolean functions of up to 12 variables, with some existing, popular, state-of-the-art methods, like EXORCISM-4 \cite{2001_MP} and EPOEM2 \cite{2016_TGPC}. We have observed that our Multi-variate Horner method is substantially faster than all the other methods. Our biclique-based algorithm achieves the largest AND-count reductions among all the methods.

\section{Preliminaries}
\label{sec:prelim}

In this paper we consider Boolean expressions in ESOP or SOP form. So a \textbf{sum} denotes either OR ($\vee$) or XOR ($\oplus$) operation, that should be clear from the context. A \textbf{product} refers to an AND ($\wedge$) operation. The number of AND (gates) in a Boolean expression is referred to as the \textbf{AND-count} of the expression. The \textbf{multiplicative complexity} of a Boolean function is the minimum number of AND (gates) that is required to implement it over the (gate) basis $\{\wedge,\oplus,1\}$ or $\{\wedge,\vee,1\}$.  

The complement of a variable, $x$, is denoted by $\conj{x} = 1\oplus x$. We say $x$ has \emph{positive polarity}, while $\conj{x}$ has \emph{negative polarity}. A \textbf{monomial} is the product of at most $n$ variables. We say that a monomial has \textbf{weight} $k$ if it is the product of $k$ variables. A \textbf{constant} is a monomial of weight 0. 
Let $\indx, \indx'\subseteq\{1,2,\ldots,n\}$. We define $m_{\indx}$ and $\conj{m}_{\indx'}$ as follows.
\begin{eqnarray}
 m_{\indx} = \prod_{j\in \indx} x_j, \qquad\text{and}\qquad \conj{m}_{\indx'} = \prod_{j\in \indx'} \conj{x}_j .
 \label{eqn:monoDefn}
\end{eqnarray}
Hence, in general, a monomial is expressed as $m_{\indx}\conj{m}_{\indx'}$, where $\indx$ and $\indx'$ are sets of subscripts of positive and negative polarity variables, respectively. We remember that the variables in a monomial commute, and so it is possible to express a monomial in the above form. A \textbf{polynomial} in $n$ variables comprises of a sum of one or more \textbf{monomials}. If the sum is XOR, then we have an ESOP (XOR Sum Of Product) expression. If the sum is OR, then we have an SOP (Sum Of Product) expression. 

\subsection{Encoding polynomial for ESOP expressions}
\label{subsec:encPoly}

In this section we mention some results that are useful in deriving the Positive Polarity Reed Muller (PPRM) expansion of a given ESOP expression, either from the expression itself or from its truth table. In a PPRM expression all variables appear in positive polarity. 

\paragraph{From the truth table :} Suppose we have an $n$-length bit string - ($b_1,b_2,\ldots, b_n$), denoted as $\vec{b}$. We encode this bit string into a polynomial in $n$ Boolean variables - $x_1, x_2,\ldots, x_n$, where variable $x_i$ corresponds to bit $b_i$. We assign the following polynomial to each variable $b_i$.
\begin{eqnarray}
 b_i &\mapsto& \frac{1+(-1)^{b_i}}{2} + x_i := p_{b_i}(x_i)
 \label{eqn:bitEncode}
\end{eqnarray}
The complete bit string $(b_1,b_2,\ldots, b_n)$ is encoded as follows.
\begin{eqnarray}
 (b_1,b_2,\ldots, b_n) \mapsto \prod_{i=1}^n\left(\frac{1+(-1)^{b_i}}{2}+x_i \right) = \prod_{i=1}^np_{b_i}(x_i) := p_{\vec{b}}(x_1,\ldots,x_n) 
 \label{eqn:stringEncode}
\end{eqnarray}
We refer to $p_{\vec{b}}(x_1,\ldots,x_n)$ as the \emph{encoding polynomial of bit string} $\vec{b}$.

\begin{lemma}[\cite{2025_Mqram}]
Suppose we have $n$ bits - $b_1,b_2,\ldots,b_n$ and we associate a variable $x_i$ to each bit $b_i$. Consider the $2^n$ encoding polynomials $\{ p_{\vec{b}}(x_1,\ldots,x_n) \}$ corresponding to the $2^n$ possible $n$-bit strings $\vec{b} = (b_1, b_2,\ldots,b_n)$, as defined in Equation \ref{eqn:stringEncode}. Then we have
\begin{eqnarray}
 p_{\vec{b}}(b_1',b_2',\ldots,b_n') \equiv \delta_{\vec{b},\vec{b'}} \mod 2,\qquad\text{where}\quad \vec{b'} = (b_1', b_2',\ldots,b_n'), \nonumber
\end{eqnarray}
implying $p_{\vec{b}}\left(b_1',b_2',\ldots,b_n'\right)\equiv 1\mod 2$ if and only if $\vec{b} = \vec{b'}$ or $b_j = b_j'$ for each $j=1,\ldots,n$. Else, it is $0 \mod 2$.
 \label{lem:uniqueVal}
\end{lemma}

\begin{corollary}
Let $f(x_1,\ldots,x_n) : \fld_2^n\rightarrow\fld_2$ be an $n$-variable function. Let $\mathcal{S}_1 = \{ \vec{b} = (b_1,\ldots, b_n) : f(\vec{b}) = 1 \}$. Then,
\begin{eqnarray}
    f(x_1,\ldots,x_n) = \bigoplus_{\vec{b}\in\mathcal{S}_1} \left( p_{\vec{b}}(x_1,\ldots,x_n) \mod 2 \right).
    \label{eqn:polyRep}
\end{eqnarray}
 \label{cor:polyRep}
\end{corollary}

The polynomial on the RHS of Eqn. \ref{eqn:polyRep} is referred to as \textbf{polynomial representation} of the function $f$. 

\begin{lemma}[\cite{2025_Mqram}]
Let $p_{\vec{b}}(x_1,\ldots,x_n)$ be the encoding polynomial corresponding to the bit string $\vec{b} = (b_1,\ldots,b_n)$, as defined in Equation \ref{eqn:stringEncode}. Assume that $k$ of the bits i.e. $b_{i_1},\ldots,b_{i_k}$ are 1 and the rest 0. Then, 
\begin{eqnarray}
 p_{\vec{b}}(x_1,\ldots,x_n) = \bigoplus_{\mathcal{I}': \mathcal{I}'\supseteq \mathcal{I} } m_{\mathcal{I}'}, \qquad\text{where}\qquad\mathcal{I} = \{ i_1,\ldots, i_k\}.   \nonumber
\end{eqnarray}
 \label{lem:monoSum}
\end{lemma}

\begin{corollary}[\cite{2025_Mqram}]
Let $p_{\vec{b}}(x_1,\ldots,x_n)$ be the encoding polynomial corresponding to the bit string $\vec{b} = (b_1,\ldots, b_n)$, as defined in Equation \ref{eqn:stringEncode}. Let $\mathcal{I}_1$ be the subset of indices of the bits in $\vec{b}$ that have value 1. Given any subset of indices $\mathcal{I}\subseteq \{1,\ldots,n\}$, the monomial $m_{\mathcal{I}}$ appears as a summand in $p_{\vec{b}}(x_1,\ldots,x_n)$ if and only if $\mathcal{I}_1 \subseteq \mathcal{I}$. 
\label{corr:monoAppear}
\end{corollary}

From the above results, we get an algorithm (PPRM-TRUTH TABLE) to derive the polynomial representation or PPRM form of a Boolean function $f$, from its truth table. 
\begin{enumerate}
    \item  $\mathcal{M}_f = \emptyset$. This set stores the monomials appearing in the polynomial representation of $f$.

    \item For each monomial $m$ (consisting of only positive polarity variables) of weight at most $n$ :
    \begin{enumerate}
        \item $q$ = Number of bit strings in $\mathcal{S}_1$ such that $m$ appears in its encoding polynomial. We use the superset relation in Corollary \ref{corr:monoAppear}, in order to check if $m$ appears in the encoding polynomial of a bit-string.

        \item If $q$ is odd then add $m$ to $\mathcal{M}_f$.
    \end{enumerate}
\end{enumerate}
The PPRM form of $f$ is the XOR of the monomials in $\mathcal{M}_f$.

\paragraph{From the function :} Suppose we are given an ESOP expression $f(x_1,\ldots,x_n)$. We can convert it to an equivalent expression consisting of only positive polarity variables by observing that for any Boolean variable $x$, we have $\conj{x} = 1\oplus x$. Hence,
\begin{eqnarray}
    m_{\indx}\conj{m}_{\indx'} &=& \left(\prod_{j\in\indx}x_j\right)\left(\prod_{k\in\indx'} (1\oplus x_k) \right) 
    = \bigoplus_{\substack{\indx''\supseteq \indx; \\ \indx''\subseteq\indx\cup\indx' }} m_{\indx''} 
    := p_{(\indx,\indx')}(x_1,\ldots,x_n).      \label{eqn:pprm_mono}
\end{eqnarray}
We refer to $p_{(\indx,\indx')}(x_1,\ldots,x_n)$ as the \emph{encoding polynomial of monomial} $m_{\indx}\conj{m}_{\indx'}$. The PPRM expansion of $f(x_1,\ldots,x_n)$ is, 
\begin{eqnarray}
    f(x_1,\ldots,x_n) &=& \bigoplus_{m_{\indx}\conj{m}_{\indx'} \in f } p_{(\indx,\indx')}(x_1,\ldots,x_n).
    \label{eqn:pprm_f}
\end{eqnarray}

Similar to Corollary \ref{corr:monoAppear}, we can infer when a monomial of positive polarity variables, appears in the encoding polynomial of a given monomial. Hence, analogous to PPRM-TRUTH TABLE, we have an algorithm to compute the PPRM expansion of a function. We refer to this procedure as PPRM-FUNCTION.

\subsection{Bicliques}
\label{subsec:biclique}

A \textbf{bipartite graph}, $G$, is a triplet $(U,V,E)$, where $U, V$ are set of vertices such that $U\cap V=\emptyset$ and $E\subseteq U\times V$ is the set of edges. In this paper we have considered undirected and unweighted graphs. A \textbf{biclique} $C = (U', V', E')$ is a complete bipartite graph, that is, for each pair of vertices $u\in U'$ and $v\in V'$, there exists an edge $e = (u,v)\in E'$. Given any graph $\hat{G} = (\hat{V},\hat{E})$, we say that $S\subseteq \hat{V}$ is a \textbf{biclique subgraph} of $\hat{G}$ if and only if the induced subgraph $\hat{G}[S]$ is a biclique. A \textbf{biclique cover} of $\hat{G}$ is a collection of set of vertices $S_1,\ldots. S_k$ such that each $\hat{G}[S_i]$ is a biclique subgraph of $\hat{G}$ and each edge $\hat{e}\in\hat{G}$ appears at least once in some $\hat{G}[S_i]$. A \textbf{biclique partition} of $\hat{G}$ is a biclique cover such that each edge is covered exactly once.

\textbf{Maximum Edge Biclique} (MEB) problem finds a biclique that maximizes the number of edges and this problem is NP-complete \cite{2002_GJ, 2003_P}. On the other hand, \textbf{Maximum Vertex Biclique} (MVB) problem finds a biclique that maximizes the number of vertices, and can be solved in polynomial time \cite{2006_LSY}. A number of algorithms have been developed to solve the MEB problem using various methods like integer programming, Monte Carlo, branch-and-bound, divide-and-conquer, etc \cite{2016_SYL, 2018_SO, 2020_LQLetal, 2025_GB}

The objective of the \textbf{Minimum Biclique Cover} (MBC) problem is to compute a collection of biclique subgraphs of $G$ that together cover all edges of $G$, while minimizing the number of such biclique subgraphs. Another related problem is the \textbf{Minimum Biclique Partition} (MBP), where the goal is to find the minimum cover such that each edge is covered by exactly one biclique. The problem of MBC is NP-complete \cite{1977_O, 1996_M}. Both MBC and MBP are NP-hard to approximate \cite{1990_S, 1993_JR, 1994_LY, 2007_GH, 2014_CHHK} and hence clever heuristic algorithms have been developed \cite{2006_SLY, 2008_EHMetal}

Given a bipartite graph $G = (U,V,E)$, a \textbf{$k$-defective biclique} $D = (U_D,V_D,E_D)$ of $G$ is a subgraph of $G$ that has at most $k$ non-edges, that is, $\left|U_D\times V_D\setminus E_D\right|\leq k$. It is NP-hard to find a connected (edge) maximum $k$-defective biclique \cite{2025_CLDetal}.

\section{Factorization of ESOP expression}
\label{sec:esopFactor}

In this section we describe two methods for factoring Boolean polynomials in ESOP form. The first method is graphical, while the second is algebraic. Both the methods work with PPRM form. So as a pre-processing step, we derive the polynomial representation or PPRM form of a given function, $f$, either from the function itself (PPRM-FUNCTION) or its truth table (PPRM-TRUTH TABLE), as described in Section \ref{subsec:encPoly}.

\paragraph{Encoding a monomial :}  Suppose the polynomials have $n$ Boolean variables - $x_0,\ldots, x_{n-1}$. We can map each monomial as an array of integers of length $n$, whose $i^{th}$ entry is $1$ if the monomial has variable $x_i$; else this entry is $0$. Thus each polynomial is stored as a set of arrays. Given $\vec{b'}, \vec{b}\in\{0,1\}^n$, a monomial $x^{\vec{b}}$ evaluates to 1 at $\vec{b'}$, if and only if the entry-wise AND of $\vec{b}$ and $\vec{b'}$ is $\vec{b}$.

Alternatively, we can map each monomial bijectively to an integer, by considering $\vec{b}$ as an array of bits, as follows. 
\begin{eqnarray}
    x^{\vec{b}} = x_{n-1}^{b_{n-1}}\ldots x_0^{b_0} = \prod_{j=n-1}^0 x_j^{b_j} \mapsto (b_{n-1},\ldots,b_0)\mapsto \sum_{j=0}^{n-1}b_j2^j
    \label{eqn:encMono}
\end{eqnarray}
Given an integer, it is straightforward to retrieve the encoded monomial. Thus each polynomial is stored as a set of integers. In this case, a monomial $x^{\vec{b}}$ evaluates to 1 for a bit string $\vec{b'}$ if and only if the bit-wise AND of $\vec{b}$ and $\vec{b'}$ is $\vec{b}$.

\begin{figure}[h]
\centering
    \begin{subfigure}[b]{0.3\textwidth}
        \centering
        \includegraphics[width=\textwidth]{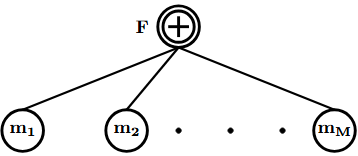}
        \caption{}
    \end{subfigure}
    \hfill
    \begin{subfigure}[b]{0.6\textwidth}
        \centering
        \includegraphics[width=\textwidth]{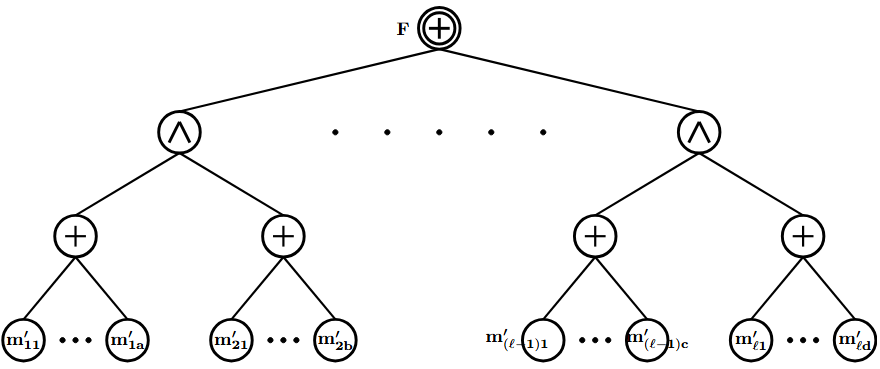}
        \caption{}
    \end{subfigure}
    \hfill
    \begin{subfigure}[b]{0.5\textwidth}
        \centering
        \includegraphics[width=\textwidth]{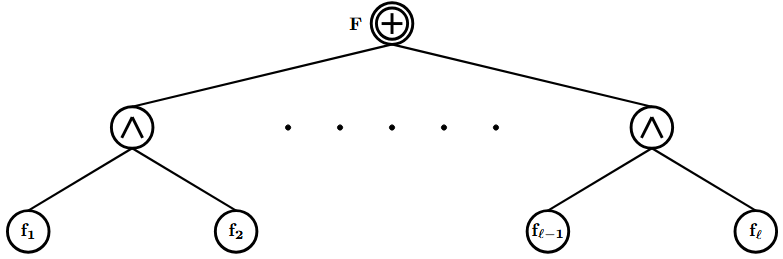}
        \caption{}
    \end{subfigure}
    \hfill
    \begin{subfigure}[b]{0.55\textwidth}
        \centering
        \includegraphics[width=\textwidth]{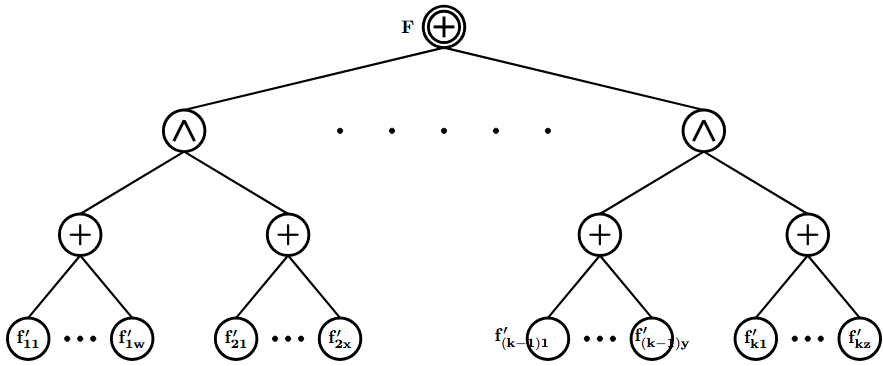}
        \caption{}
    \end{subfigure}
    \hfill 
    \begin{subfigure}[b]{0.5\textwidth}
        \centering
        \includegraphics[width=\textwidth]{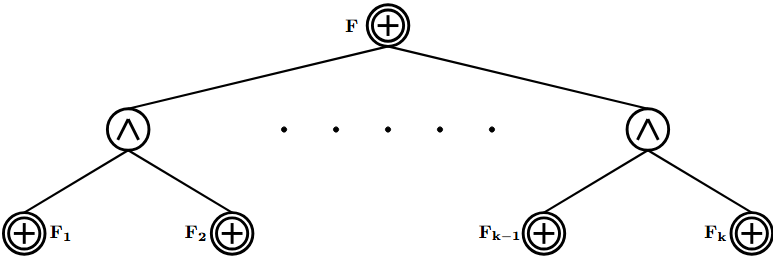}
        \caption{}
    \end{subfigure}
    \caption{One iteration of ESOP-FACTORING. (a) A simple ESOP expression, $F = \bigoplus_{i=1}^Mm_i$, represented by a tree whose root (level 0) is a XOR-node and the leaves are the summand monomials. (b) The tree, after ESOP-FACTOR-TYPE-I. The children of each XOR node at level 2, form a factor. (c) The equivalent tree of (b), where each sub-tree  rooted at level-2 XOR node, is replaced by equivalent expression. (d) The tree, after applying ESOP-FACTOR-TYPE-II. Again, the leaves of each XOR-subtree form a factor. (e) The equivalent tree of (d), where each sub-tree below the AND-nodes, is replaced by equivalent function. Each of these leaf nodes become roots for the next iteration.   }
    \label{fig:esop_biclique}
\end{figure}

\subsection{ESOP factorization using bicliques}
\label{subsec:esopBiclique}

Our ESOP factoring algorithm consists of a number of iterations of two types of factoring.
\begin{enumerate}
    \item Factoring an expression or function, consisting of XOR of monomials. These expressions or functions are referred to as \emph{simple}.

    \item Factoring a \emph{complex} expression or function, which itself is XOR of a number of product expressions. \emph{Product expressions} are those that are the product of a number of expressions.
\end{enumerate}
Though the type-I factoring is a special case of the type-II factoring, but we consider them separately. The reason will be clear from the later explanations.

\subsubsection*{Type-I factoring : Factoring simple expressions}

We first define the following terms.

\begin{definition}[\textbf{Factor and co-factor of a monomial}]
Let $m = \prod_{j\in\indx} x_j$ and $m' = \prod_{k\in\indx'}x_k$ be two monomials, where $\indx, \indx'$ index the sets of variables that appear in these monomials. $m'$ is a \textbf{factor} of $m$ if and only if $\indx' \subseteq\indx$. If $\indx' = \indx$ (i.e. $m' = m$) or $\indx' = \emptyset$ (i.e. $m' = 1$) then $m'$ is referred to as a \textbf{trivial factor}.

The monomial $m'' = m/m' = \prod_{\ell\in\indx\setminus\indx'} x_{\ell} $ is referred to as the \textbf{co-factor of $m$ with respect to $m'$}.  
\label{defn:factorMono}
\end{definition}

\begin{definition}[\textbf{Largest Common Factor (LCF)}]
Given a set of monomials, $ \mathcal{M} = \{m_1, m_2, \ldots, m_k\}$, a monomial $m$ is referred to as the Largest Common Factor of this set, if and only if (i) $m$ is a (common) factor of each monomial $m_j\in\mathcal{M}$; (ii) there does not exist any other monomial $m'$ such that $m'$ is also a common factor of each monomial in $\mathcal{M}$ and $m$ itself is a factor of $m'$.

Alternatively, we can state as follows. Suppose $\mathcal{M} = \{ m_j = \prod_{i\in\indx_j} x_i : j = 1,\ldots,k \}$ be a set of monomials. $m = \prod_{i\in\indx} x_i$ is the LCF of this set of monomials if and only if (i) $\indx\subseteq\indx_j$ for each $j = 1,\ldots,k$; (ii) there does not exist any set $\indx'\neq\indx$ such that $\indx\subset\indx'\subseteq\indx_j$ for each $j$.
\label{defn:LCF}
\end{definition}

\paragraph{Procedure to find pair-wise LCF and corresponding co-factors : } Suppose there are M monomials. We want to list the pairwise LCFs and their cofactors. The LCF of the monomials $m_1$ and $m_2$ is exactly equal to the entry-wise AND of the array of integers/bits that encode these monomials. The time complexity is $O(nM^2)$. Given a monomial $m_1$ and a factor $m'$, the corresponding cofactor $m_1/m'$ can be obtained by computing the entry-wise XOR of the arrays.

\begin{definition}[\textbf{Factor-Cofactor Bipartite graph of simple expression}]
Suppose $f$ is a simple function, that is expressed as a sum of monomials. Let $\mathcal{M}$ be the set of summand monomials of $f$. $U = \{ \text{LCF}(m_i,m_j) : m_i, m_j\in\mathcal{M} \}$ is a set of pairwise LCFs of the monomial summands of $f$. $V = \{m/m_u : m\in\mathcal{M},\quad m_u\in\mathcal{U}   \}$ is a set of the cofactors of the summand monomials with respect to the LCFs. Then, the factor-cofactor bipartite graph of $f$ is $G_f=(U,V,E)$, where an edge $(u,v)$ is in $E$ if and only if $uv\in\mathcal{M}$.
    \label{defn:factCofactGraph}
\end{definition}

Suppose $G' = (U', V', E')$ is a biclique in $G_f$. Let $U' = \{u_1',u_2',\ldots,u_k'\}$ and $V' = \{v_1',v_2',\ldots,v_{\ell}'\}$ be the sets of factors and co-factors, respectively. Also, $E' = \{e_1',e_2',\ldots, e_{k\ell}'\}$ be the set of summand monomials that appear as edges in this biclique. Then, by the above construction, we have
\begin{eqnarray}
    \left(\bigoplus_{i=1}^ku_i'\right) \left(\bigoplus_{j=1}^{\ell} v_j'\right) = \bigoplus_{i=1}^{k\ell}e_i'.  \nonumber
\end{eqnarray}
So, a biclique in $G_f$ identifies a pair of factors of $f$. Let $\mathcal{S}_{G_f} = \{G_1' = (U_1',V_1',E_1'),\ldots,G_b'=(U_b',V_b',E_b') \}$ be a set of bicliques in $G_f$, such that each edge in $G_f$ is covered by an odd number of bicliques in $\mathcal{S}_{G_f}$. We say that a monomial is covered, if the edge(s) corresponding to this monomial is (are) covered. A factorization of $f$ is as follows.
\begin{eqnarray}
    f = \bigoplus_{i=1}^b \left( \left(\bigoplus_{u'\in U_i'}u'\right) \left(\bigoplus_{v'\in V_i'} v'\right) \right)
    \label{eqn:factorSimpleF}
\end{eqnarray}
Hence, factoring a simple expression with the minimum number of factors, is akin to the problem of Minimum Biclique Cover (MBC), but with the following differences.
\begin{enumerate}
    \item There are two edges for each monomial in $f$.
    
    \item Each edge, corresponding to monomials in $f$, is covered by an odd number of bicliques.

    \item We can add edges to $G_f$, corresponding to monomials that are not in $f$. These extra edges must be covered by an even number of bicliques. 
\end{enumerate}
Now, we describe a procedure that takes as input an ESOP function $F$, in PPRM form. $F$ is stored as a set of monomials. The goal is to return a number of factored expressions $f_1f_2, f_3f_4$, etc, such that
\begin{eqnarray}
    F = \left(\bigoplus_i f_{2i-1}f_{2i}\right) \oplus f_K
\end{eqnarray}
Each of the factored expression, $f_{2i-i}f_{2i}$, corresponds to a biclique in $G_F$, the Factor-Cofactor Bipartite graph of $F$. So, we perform a number of iterations and in each iteration we find a biclique. Now, condition (2) says we can cover all edges odd number of times; while due to condition (3), we are allowed to add edges, but these have to be covered even number of times. So, we can search for defective bicliques. For optimizing the number of factors, we can also search for maximum edge biclique, or a maximum $k$-defective biclique, where $k$ is a parameter determined at input. We remember that for $k=0$, a defective biclique is a non-defective one. We maintain a list, $\mathcal{A}$, of monomials that have been uncovered so far. Here we mention that for simplicity, in $G_F$ we keep only one edge corresponding to each monomial (condition (1)). Keeping both edges will increase the number of possible biclique covers, ane hence may lead to better optimizations. But it will also increase the computational complexity. So we randomly keep any one of the edges, for each monomial. 

After we find a biclique (defective or not), we store the corresponding factors. For each edge, $e$, in this biclique, let $m_e$ be the corresponding monomial. This monomial can be obtained by multiplying the monomials corresponding to the two end vertices of this edge. There can be two possibilities. First, $m_e$ is in $\mathcal{A}$. In this case, we remove the edge in $G_F$, that corresponds to $m_e$. We also, remove $m_e$ from $\mathcal{A}$. Second, $m_e\notin \mathcal{A}$. In this case, we add the edge corresponding to $m_e$ to $G_F$. We also add $m_e$ to $\mathcal{A}$. In this way we ensure that conditions (2) and (3) regarding odd and even overlaps, are satisfied. If we work with non-defective bicliques then we are only allowed to add edges that are in $E$ before the start of all iterations. 

In the next iteration, we find a biclique in the modified bipartite graph and repeat the same procedure. We keep iterating till $\mathcal{A}$ becomes an empty set. We call this procedure ESOP-FACTOR-TYPE-I and a pseudocode has been given in Algorithm \ref{alg:esp-type-I}.

 \begin{algorithm}
     \scriptsize
     \caption{ESOP-FACTOR-TYPE-I}
     \label{alg:esp-type-I}

    \KwIn{ESOP function $F$ in PPRM form, given as a set of monomials}

    \KwOut{A set, $\mathcal{B}$, of factored expressions}

    Construct the Factor-Cofactor Bipartite Graph, $G_F = (U, V, E)$.  \;

    $\mathcal{A}\leftarrow $ Set of monomials in $F$    \;

    $\mathcal{B}\leftarrow\emptyset$     \;

    \While{$\mathcal{A}\neq\emptyset$}
    {
        Find a (defective/quasi/maximum edge) biclique $G' = (U',V',E')$, in $G_F$ \;

        Add the corresponding factors to $\mathcal{B}$  \;

        \For{each $e\in E'$}
        {
            $m_e\leftarrow $ Monomial corresponding to $e$  \;

            \eIf{$m_e\in\mathcal{A}$}
            {
                $E_{m_e}\leftarrow $ Set of edges in $G_F$, whose edges correspond to monomial $m_e$     \;

                $E\leftarrow E\setminus E_{m_e}$    \;

                $\mathcal{A}\leftarrow\mathcal{A}\setminus\{m_e\}$  \;
            }
            {
                $\mathcal{A}\leftarrow\mathcal{A}\cup\{m_e\}$  \;

                Add both the edges that correspond to $m_e$, to $E$ \;
            }
        }
    }
     
 \end{algorithm}

\subsubsection*{Type-II factoring : Factoring complex expressions}

Here we consider complex expressions, that is, those which are XOR of other expressions. Such a complex expression can be represented by a set of a number of expressions or functions. We define the following.

\begin{definition}[\textbf{Factor and cofactor of function/expression}]
Given an expression $F$, we say that another function $F'$ is a factor of $F$ if and only if $F$ can be expressed as $F=F'\wedge F''$, where $F''$ is another expression. $F''$ is called the cofactor of $F$ with respect to $F'$. If $F' = F$ or 1 then we have a trivial factor.
    \label{defn:factorFn}
\end{definition}

\begin{definition}[\textbf{Largest Common Factor of a set of expressions}]
Suppose we have a set of expressions $\mathcal{F} = \{F_1,\ldots, F_k\}$. An expression $F'$ is the LCF of $\mathcal{F}$ if and only if (i) $F'$ is a (common factor) of each $F_j\in\mathcal{F}$; (ii) there does not exist any expression $F''$ such that $F'$ is a factor of $F''$, which itself is a factor of each $F_j$.
    \label{defn:LCFfn}
\end{definition}
An expression is stored either as a set of array of integers or set of integers encoding the monomials. So, the LCF of a pair of expressions is the set of arrays or integers that belong to both the sets. 
Similar to Type-I factoring, we can construct a \textbf{Factor-Cofactor Bipartite graph of complex expression}, except this time each vertex is a LCF or cofactor of a set of expressions and each edge is an expression. The procedure for Type-II factoring is similar to ESOP-FACTOR-TYPE-I, except we incorporate the above mentioned modifications for expressions. We refer to this procedure as \textbf{ESOP-FACTOR-TYPE-II}.

\subsubsection*{Overall procedure : ESOP-FACTORING} 

In the ESOP-FACTORING algorithm, we perform these two types of factoring alternatively. A pseudocode has been given in Algorithm \ref{alg:esopFactor}. The input of this algorithm is a function $F$ given as a set of monomials. First we factor $F$, that is a simple expression, using ESOP-FACTOR-TYPE-I procedure. Thus after this procedure we can write
\begin{eqnarray}
    F = \left(\bigoplus_i f_{2i-1}f_{2i} \right) \oplus f_{\ell+1};  \label{eqn:F_typeI}
\end{eqnarray}
where $f_{\ell+1}$ is a simple expression and it cannot be factored further. Next, we use ESOP-FACTOR-TYPE-II to factor the complex expressions. For example, if we have terms like $f_if_j \oplus f_if_{\ell}$, then we can extract a common factor and write $f_i(f_j\oplus f_{\ell})$. Thus, after this step,
\begin{eqnarray}
    F = \left(\bigoplus_i F_{2i-1}F_{2i} \right) \oplus F_{k+1};  \label{eqn:F_typeII}
\end{eqnarray}
where each factor $F_i$ is a XOR of monomials, and hence a simple expression. Then we recursively call ESOP-FACTORING algorithm with each $F_i$. If $F_{k+1} = f_{\ell+1}$ then we do not factor $F_{k+1}$, since it implies it cannot be factored.

\begin{algorithm}
    \scriptsize
    \caption{ESOP-FACTORING}
    \label{alg:esopFactor}

    \KwIn{ $F = \{ m_1,m_2,\ldots,m_M \}$ }

    \KwOut{ A factorization of $F$. }

    \If{LCF of each pair of monomials is 1 }
    {
        \textbf{return} $F$ \;
    }

    $\{f_1f_2, f_3f_4,\ldots, f_{\ell-1}f_{\ell}, f_{\ell+1} \} \leftarrow $ ESOP-FACTOR-TYPE-I$(m_1,m_2,\ldots,m_M)$   \;

    $\{F_1F_2, F_3F_4, \ldots, F_{k-1}F_{k}, F_{k+1} \}\leftarrow$ ESOP-FACTOR-TYPE-II$(f_1f_2, f_3f_4,\ldots, f_{\ell+1} )$  \;

    \For{$i=1,\ldots,k$}
    {
           ESOP-FACTORING$(F_i)$    \;
    }

    \If{$f_{\ell+1} != F_{k+1}$}
    {
        ESOP-FACTORING$(F_{k+1})$        \;
    }
    SIMPLIFY$(F)$
\end{algorithm}

\begin{algorithm}
    \scriptsize
    \caption{SIMPLIFY}
    \label{alg:simplify_meth}

    \KwIn{ Some expression $E$}

    \KwOut{ A simplified version of $E$ }

    \If{$E = AND(E_1, ..., E_l)$ and $l = 1$ }
    {
        \textbf{return} SIMPLIFY$(E_1)$ \;
    }
  
    \If{$E = AND(E_1, ..., E_l)$} {
          $F\leftarrow [\textrm{SIMPLIFY}(E_i), s.t. E_i \neq True]$\;
          
         \textbf{return} $AND(DEDUP(F))$ \; // DEDUP deduplicates the list of children $F$
    }
    \If{$E = XOR(E_1, ..., E_l)$}{
         \textbf{return} $XOR\left(\textrm{SIMPLIFY}(E_1), ..., \textrm{SIMPLIFY}(E_l)\right)$ \;
    }
\end{algorithm}

We can represent a Boolean expression by a tree, where each node is labeled by a monomial or function or operator (AND, XOR or OR). The children of an operator node are composed with the operator in the node (parent). For example, if the children of a XOR node are monomials $m_1, m_2, m_3$, then the sub-tree rooted at the XOR node represents a function $m_1\oplus m_2\oplus m_3$. In order to obtain the function represented by a tree, we start with the sub-trees rooted at the parents of the leaves. We replace these sub-trees with equivalent expressions and keep iterating, till we have a single node. We say that the root is at level 0 and the remaining nodes are at higher levels. 

In Figure \ref{fig:esop_biclique} we have described one iteration of ESOP-FACTORING algorithm with a similar tree. Initially, the input simple expression, $F = \bigoplus_{i=1}^Mm_i$, is represented by a tree, whose root is a XOR-node and the leaves are the summand monomials (Figure \ref{fig:esop_biclique}(a)). After applying ESOP-FACTOR-TYPE-I procedure we obtain the tree in Figure \ref{fig:esop_biclique}(b), where each sub-tree rooted at the XOR-parent-nodes of the leaves is a factor. Then, the expression is
\begin{eqnarray}
F = \left(\left(m_{11}'\oplus\cdots\oplus m_{1a}'\right)\wedge\left(m_{21}'\oplus\cdots\oplus m_{2b}'\right)\right)\oplus\cdots\oplus\left(\left(m_{(\ell-1)1}'\oplus\cdots\oplus m_{(\ell-1)c}'\right)\wedge\left(m_{\ell 1}'\oplus\cdots\oplus m_{\ell d}'\right)\right).    \nonumber
\end{eqnarray}
 In Figure \ref{fig:esop_biclique}(c) we replace each of these sub-trees with a node, labeled by the equivalent simple expression, i.e. $\bigoplus_im_{ji} = f_j$. Hence, the expression is
\begin{eqnarray}
    F = f_1f_2\oplus\cdots\oplus f_{\ell-1}f_{\ell}.    \nonumber
\end{eqnarray}
For simplicity, we have assumed that there is no part that cannot be factored, that is, $f_{\ell+1}$ in step 4 of Algorithm \ref{alg:esopFactor} is $0$. Next, we apply ESOP-FACTOR-TYPE-II and obtain the tree in Figure \ref{fig:esop_biclique}(d), where each sub-tree rooted at the XOR-nodes of level 2, is a factor. So, the expression is
\begin{eqnarray}
F = \left(\left(f_{11}'\oplus\cdots\oplus f_{1w}'\right)\wedge\left(f_{21}'\oplus\cdots\oplus f_{2x}'\right)\right)\oplus\cdots\oplus\left(\left(f_{(k-1)1}'\oplus\cdots\oplus f_{(k-1)y}'\right)\wedge\left(f_{k 1}'\oplus\cdots\oplus f_{k z}'\right)\right).    \nonumber
\end{eqnarray}
Again, we can replace each of these sub-trees with an equivalent simple expression, represented by a node (Figure \ref{fig:esop_biclique}(e)). That is,
\begin{eqnarray}
    F = F_1F_2\oplus\cdots\oplus F_{k-1}F_k.    \nonumber
\end{eqnarray}
Each of these nodes is a XOR-node and hence we factor them in the next iterations. That is, each of these become the root node and the same procedure is repeated. 
The final expression is simplified before being returned using a simple recursive method. The SIMPLIFY method essentially de-duplicates children of AND nodes and removes trivial children.

\subsection{ESOP factorization via multivariate Horner method}
\label{subsec:esopHorner}

Horner's algorithm for polynomial factorization is a straightforward method that consists in recursively rewriting a polynomial $P$ over a single variable $x$ as :$$ P(x) = A + xB(x) $$ where $A$ is a constant and $B(x)$ is another polynomial. The method is then called again on $B$, up until $B$ is constant. The algorithm is deterministic and normalizes the polynomial in a factorization that relies on $O(d)$ multiplications.

In the case of multivariate polynomials \cite{pena2000multivariate}, we can generalize this method by picking some variable $x_i$ and factorizing some polynomial $P(x_1, ..., x_n)$ as $$P(x_1, ..., x_n) = A(x_1, ..., x_{i-1}, x_{i+1}, ..., x_n) + x_i B(x_1, ..., x_{i-1}, x_{i+1}, ..., x_n) $$

Let $\operatorname{ANDS}(P)$ be the number of naive multiplications required to evaluate some polynomial $P$. We can easily see that the previous expression will use $\operatorname{ANDS}(A) + \operatorname{ANDS}(B) + 1$ multiplications.
In contrast, if we do not factor out $x_i$ from the monomials in $B$, then we would require $\operatorname{ANDS}(A) +\operatorname{ANDS}(B) + M_B$ multiplications, where $M_B$ is the number of monomials in $B$. Thus we save up to $M_B - 1$ multiplications.

This leads us to adopt the following greedy (recursive) method:
\begin{itemize}
    \item for each variable $x_i$, count how many monomial in $P$ contain $x_i$
    \item select the variable which maximizes this count
    \item split $P$ into the two corresponding polynomial $A$ and $B$ and recursively call the method on $A$ and $B$
\end{itemize}

Algorithm \ref{alg:multivarhorner} provides the pseudo code corresponding to this greedy heuristic.

\SetKwFunction{FnMH}{MultivarHorner}%
\SetKw{Return}{Return}
\begin{algorithm}
\caption{Multivariate-Horner}\label{alg:multivarhorner}
\FnMH{$P$}{

\KwData{Some polynomial $P(x_1, ..., x_n)$ as a sum of monomials $M=\{ m_1, ..., m_k\}$}
\KwResult{A factorization of $P$}
$s = \max_{i\in [n]}|\{ m\in M, x_s \in m\}|$

$A = \{m \in M, x_s \notin m\}$

$B = \{m\setminus\{x_s\}, m \in M, x_s \in m\}$

\Return{$\FnMH{A} + x_s \FnMH{B}$}
}
\end{algorithm}

\section{Factorization of SOP expression}
\label{sec:sopFactor}

In this section we describe an algorithm for factoring Boolean SOP polynomials. The algorithm uses bicliques, and broadly, is similar to ESOP-FACTORING (Algorithm \ref{alg:esopFactor}). Due to the functional differences between the XOR and OR operator, some modifications are required before Algorithm \ref{alg:esopFactor} can be applied to factoring of SOP expressions. First we state these differences and later we describe the modifications.

\begin{enumerate}
    \item Using $1\oplus x = \conj{x}$, we can derive a PPRM expansion of an ESOP expression (see Section \ref{subsec:encPoly}). But, $1\vee x = 1$, and so the results in Section \ref{subsec:encPoly} do not apply. Hence, our SOP polynomials have variables of both positive and negative polarity. This implies we cannot use an array of $\{0,1\}$ or Equation \ref{eqn:encMono} in order to encode monomials.

    \item Since $x\oplus x = 0$, in Type-I and II factoring (Section \ref{subsec:biclique}) we cover each monomial of an ESOP expression by an odd number of bicliques. Monomials not in the expression, are covered by an even number of bicliques. But $x \vee x = x$, and so we cannot add monomials to an SOP expression. The monomials already within the expression has to be covered at least once.    
\end{enumerate}

First, we describe an encoding that includes negative polarity variables.
\paragraph{Encoding a monomial :} Suppose an SOP polynomial involves $n$ variables ($x_{n-1}, \ldots, x_0$). We encode each monomial by an array $\vec{b}$ of length $2n$.    
\begin{enumerate}
    \item If $x_j$ appears in the monomial then $\vec{b}[2j+1]\vec{b}[2j] = 10$. 

    \item If $\conj{x_j}$ appears in the monomial then $\vec{b}[2j+1]\vec{b}[2j] = 01$.

    \item If $x_j$ does not appear in the monomial then $\vec{b}[2j+1]\vec{b}[2j] = 00$. 
\end{enumerate}

Alternatively, we can store $\vec{b}$ as an array of bits. More specifically, we store the integer corresponding to this array of bits. For example, we store $x_3\conj{x_1}$ as $10000100 = 2^7+2^2 = 132$. We observe that this is a bijective mapping and it is straightforward to retrieve the monomial from the encoding integer. 

\paragraph{Encoding a binary string :}  If a Boolean function is input as a truth table, then it can be specified by the set of binary strings that evaluates to 1 in the truth table. In this case we can encode each binary string either as a $2n$-length array of integers ($0$ or $1$); or bits. Specifically, an $n$-bit binary string ($\vec{a} = (a_{n-1}\ldots a_0)$) can be encoded as an array, $\vec{a}_{enc}$, of length $2n$, as follows. 
\begin{enumerate}
    \item If $a_j = 0$, then $\vec{a}_{enc}[2j+1]\vec{a}_{enc}[2j] = 01$. 

    \item If $a_j = 1$, then $\vec{a}_{enc}[2j+1]\vec{a}_{enc}[2j] = 10$. 
\end{enumerate}
Alternatively, we can consider $\vec{a}_{enc}$ as an array of bits, and store the corresponding integer. For example, we encode $1100$ as $10100101 = 165$. 

It is straightforward to evaluate a monomial $m_{\indx}\conj{m}_{\indx'}$ at a binary string $\vec{t}$. Here $\indx,\indx'\subseteq\{0,\ldots,n-1\}$ and $\indx\cap\indx'=\emptyset$. Compute the arrays or integers $\vec{a}_{mono}$ and $\vec{a}_{int}$, encoding $m_{\indx}\conj{m}_{\indx'}$ and $\vec{t}$, respectively. Compute the entry-wise or bit-wise AND, $temp$. If $temp == \vec{a}_{mono}$ then the monomial $m_{\indx}\conj{m}_{\indx'}$ evaluates to 1, else it is 0.

\subsection{SOP factorization using bicliques}
\label{subsec:sopBiclique}

As in the case of ESOP expressions, our main algorithm consists of a number of iterations of two types of factoring - (a) type-I, where a \emph{simple} expression is factored; (b) type-II, where a \emph{complex} expression is factored. The definitions of simple and complex expression, factor and co-factor, Largest Common Factor (LCF), factor-cofactor bipartite graph, are same as in Section \ref{subsec:esopBiclique}, except the fact that the sum is OR. The pair-wise LCF of a set of monomials can be obtained by computing the entry-wise or bit-wise AND of the encoding arrays or integers (respectively) of pairs of moomials, and the co-factors can be obtained by computing the entry-wise or bit-wise XOR, as done in the case of ESOP expressions. The procedure to find the pair-wise LCF of a set of expressions is also similar. 

The arguments leading to Equation \ref{eqn:factorSimpleF}, hold in this case as well, except the fact that the sum is OR in this case and hence, the XOR in this equation is replaced by OR. So, factoring a simple expression reduces to the problem of Minimum Biclique Cover (MBC). The differences 2 and 3, listed after Equation \ref{eqn:factorSimpleF} in Section \ref{subsec:esopBiclique} do not apply here, because $x\vee x = x$. Neither can we add extra edges, nor are there any restrictions on the number of times each edge (corresponding to an existing monomial) need to be covered.
The algorithm stops when each edge is covered at least once.
We also adapt the SIMPLIFY method (Alg. \ref{alg:simplify_meth}) to account for the change from XOR to OR.

\section{Implementation and benchmarks}
We implemented our algorithm\footnote{https://github.ibm.com/ibm-q-research/factorust} in Rust and benchmarked it on sets of random truth
tables with varying sparsity. We compared our factoring method against the
following baselines:

\begin{itemize}
    \item The raw truth table implementation (\textbf{initial}): each non-zero
        entry is implemented as the \textsc{and} of the appropriate polarized
        input variables. The \textsc{and} count depends only on the number of
        variables~$n$ and the number of non-zero entries~$p$, giving a total of
        $p(n-1)$ \textsc{and} gates.

    \item The naive polynomial evaluation obtained after running the polynomial
        encoding (\textbf{polynomial}): each monomial is implemented by
        computing the \textsc{and} of all variables appearing in it.

    \item The \textsc{epoem2} algorithm (\textbf{epoem2}), from~\cite{2016_TGPC},
        re-implemented in Rust.

    \item The \textsc{exorcism-4} algorithm (\textbf{exorcism\_4}),
        from~\cite{2001_MP}, re-implemented in Rust.

    \item The multivariate Horner method introduced in this work
        (\textbf{horner}).
\end{itemize}

To this list we add our two new methods:

\begin{itemize}
    \item Our factoring algorithm using the branch-and-bound max-biclique solver
        from~\cite{2020_LQLetal} (\textbf{biclique\_max}).

    \item The same factoring algorithm using the greedy large-biclique heuristic
        of~\cite{2025_CLDetal} (Section~4.3) (\textbf{biclique\_large}).
\end{itemize}

In both cases, the bipartite graph is greedily decomposed by iteratively
extracting a biclique (via the exact or heuristic method), removing it from the
graph, and repeating until the graph is empty.

\begin{figure}[h]
    \centering
    \begin{subfigure}{0.48\textwidth}
        \includegraphics[width=\textwidth]{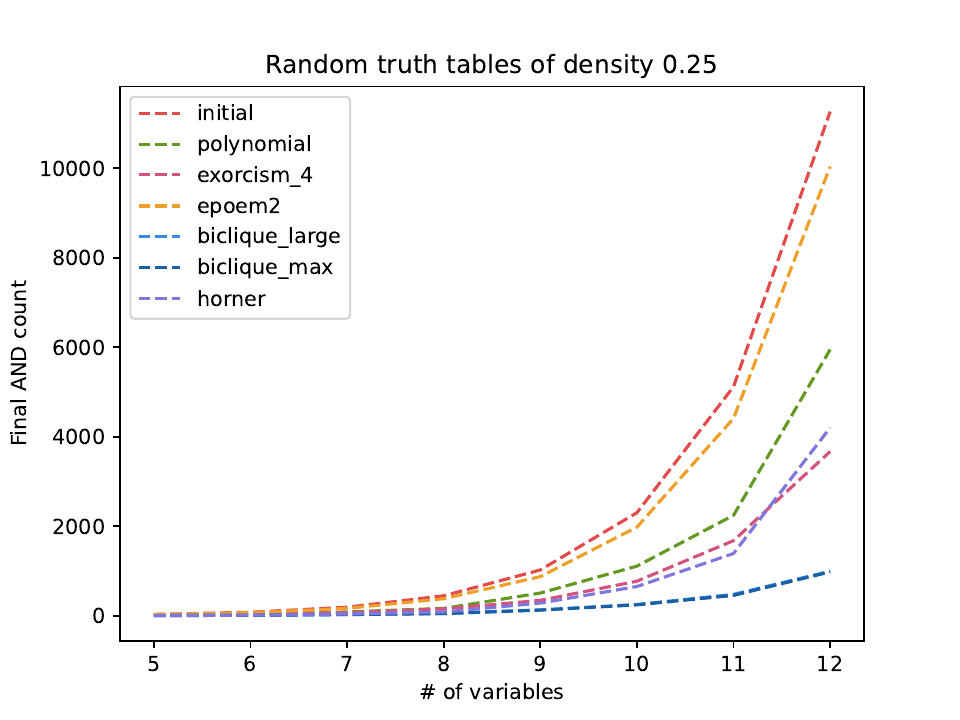}
        \caption{}
    \end{subfigure}
    \hfill
    \begin{subfigure}{0.48\textwidth}
        \includegraphics[width=\textwidth]{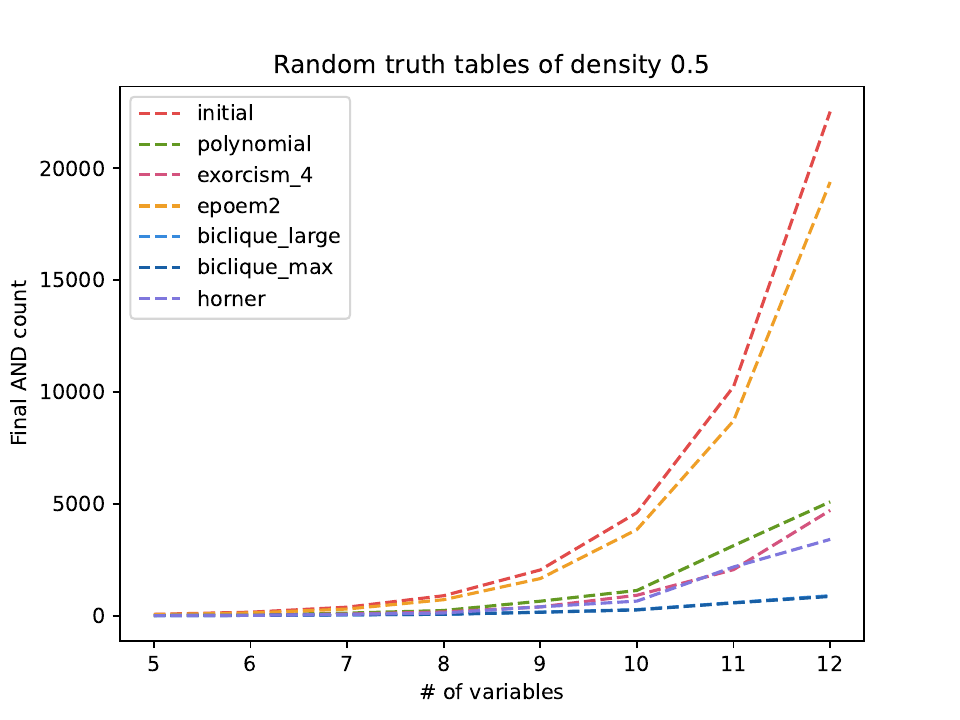}
        \caption{}
    \end{subfigure}
    \hfill
    \begin{subfigure}{0.48\textwidth}
        \includegraphics[width=\textwidth]{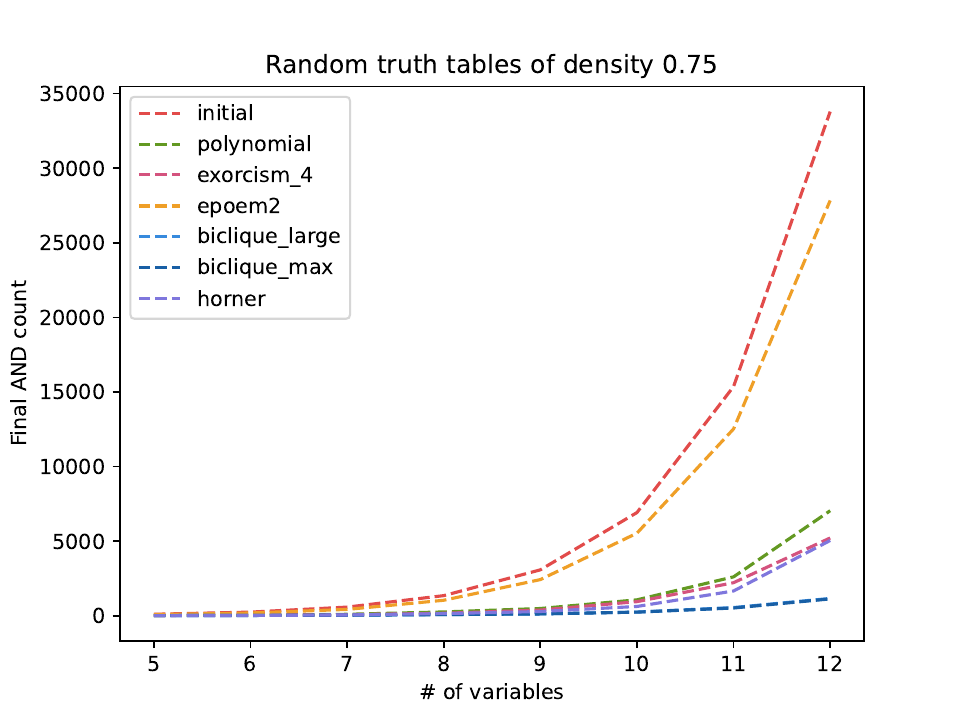}
        \caption{}
    \end{subfigure}
    \hfill
    \begin{subfigure}{0.48\textwidth}
        \includegraphics[width=\textwidth]{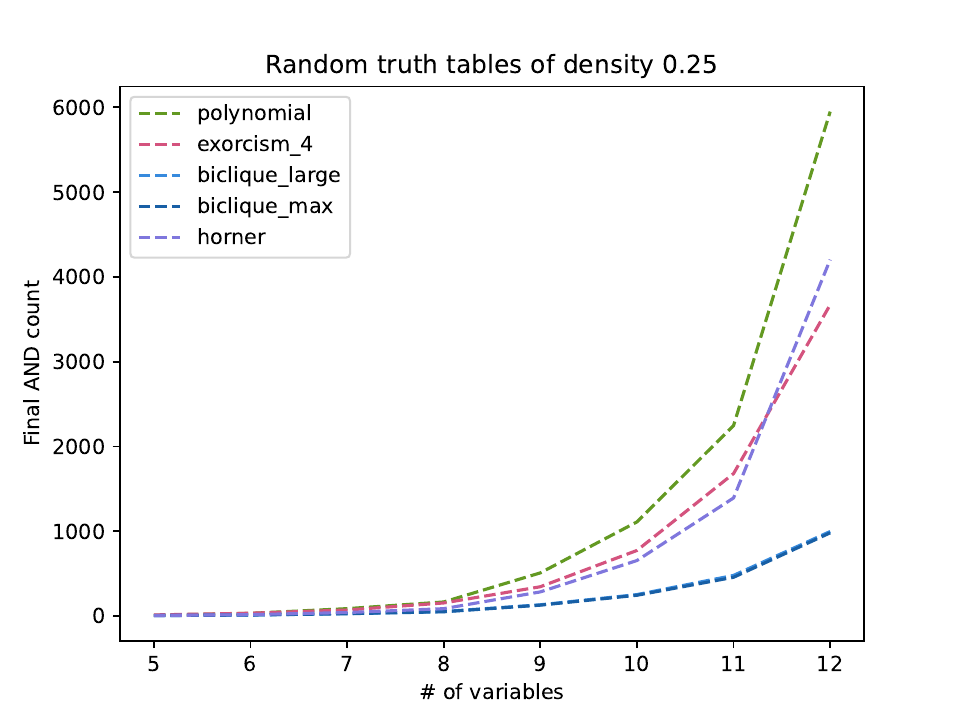}
        \caption{}
    \end{subfigure}
    \hfill
    \begin{subfigure}{0.48\textwidth}
        \includegraphics[width=\textwidth]{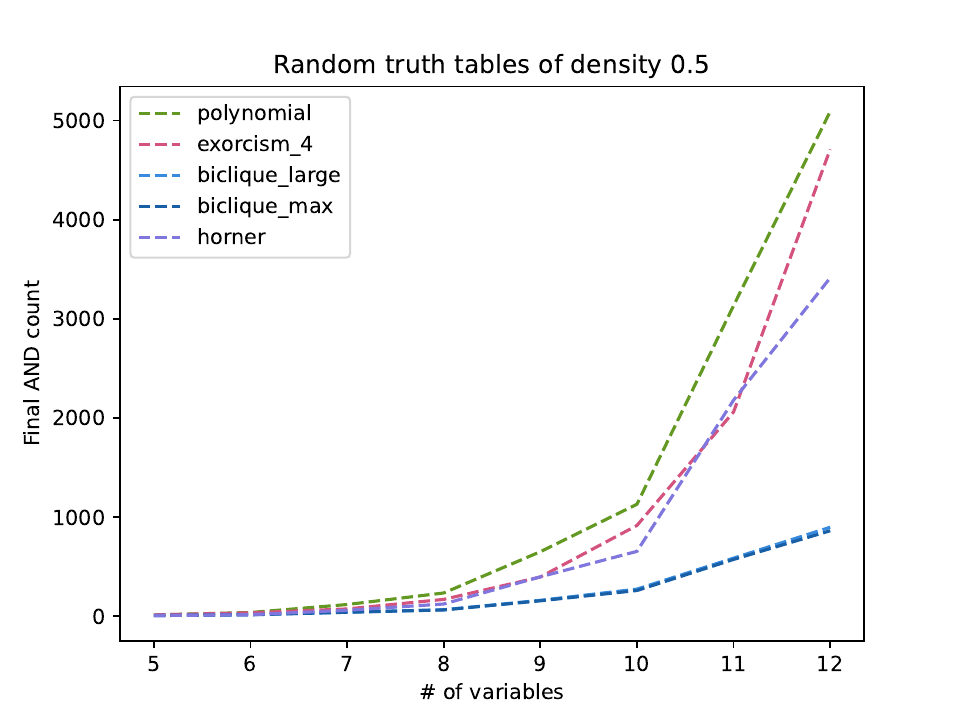}
        \caption{}
    \end{subfigure}
    \hfill
    \begin{subfigure}{0.48\textwidth}
        \includegraphics[width=\textwidth]{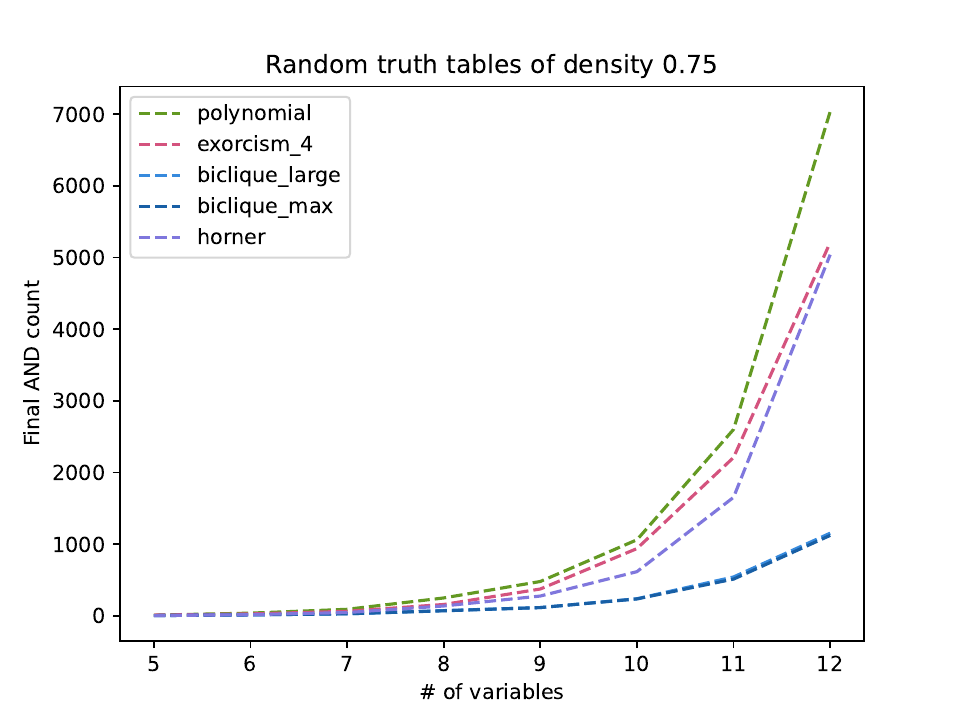}
        \caption{}
    \end{subfigure}
    \caption{AND counts of the final expressions obtained by running different heuristics on random truth tables. Each point is averaged over 10 random truth tables.
    Plots (d), (e), (f) exclude the initial AND count for a naive implementation of the truth table, and the AND count obtained by running the {\bf epoem2} heuristic.}
    \label{fig:and_count_random}
\end{figure}

\begin{figure}[h]
\centering
    \begin{subfigure}{0.48\textwidth}
    \includegraphics[width=\textwidth]{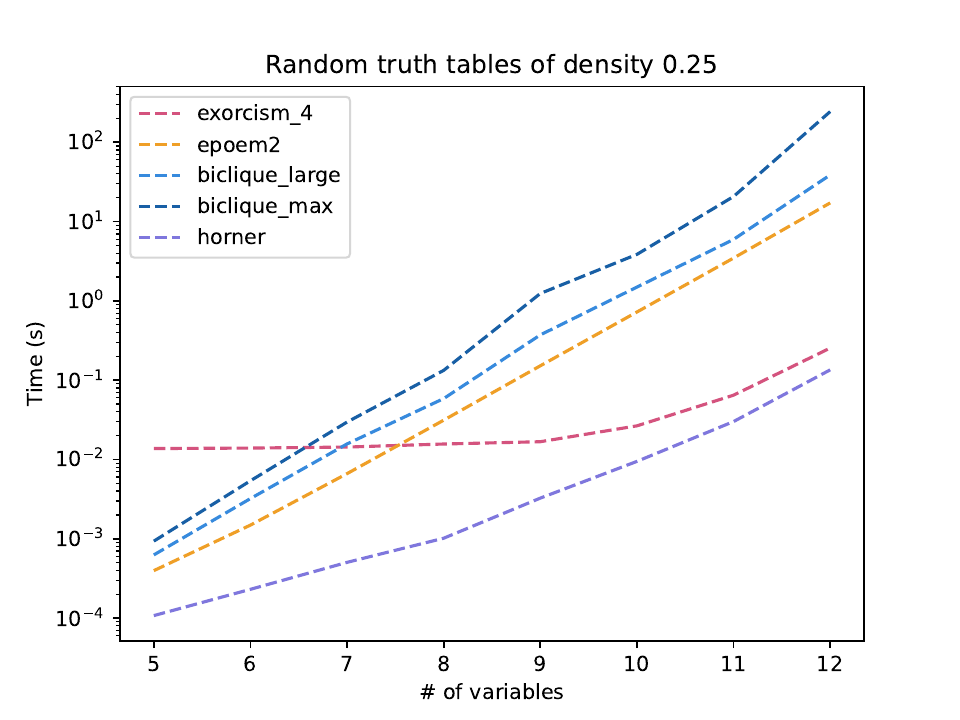}
    \caption{}
    \end{subfigure}
    \hfill
    \begin{subfigure}{0.48\textwidth}
        \includegraphics[width=\textwidth]{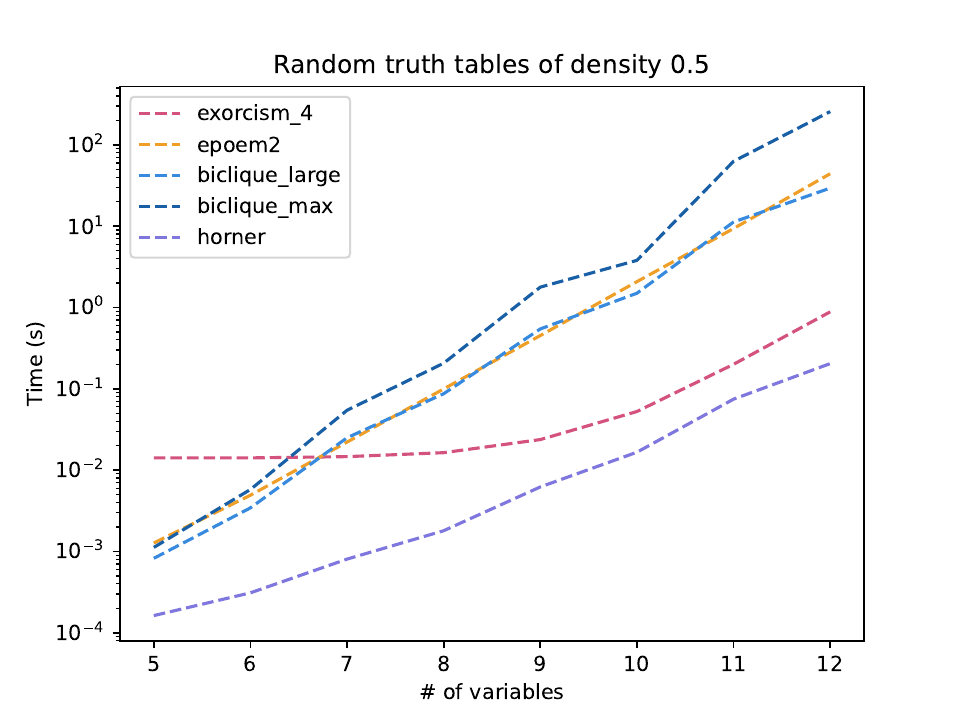}
        \caption{}
    \end{subfigure}
    \hfill
    \begin{subfigure}{0.48\textwidth}
        \centering
        \includegraphics[width=\textwidth]{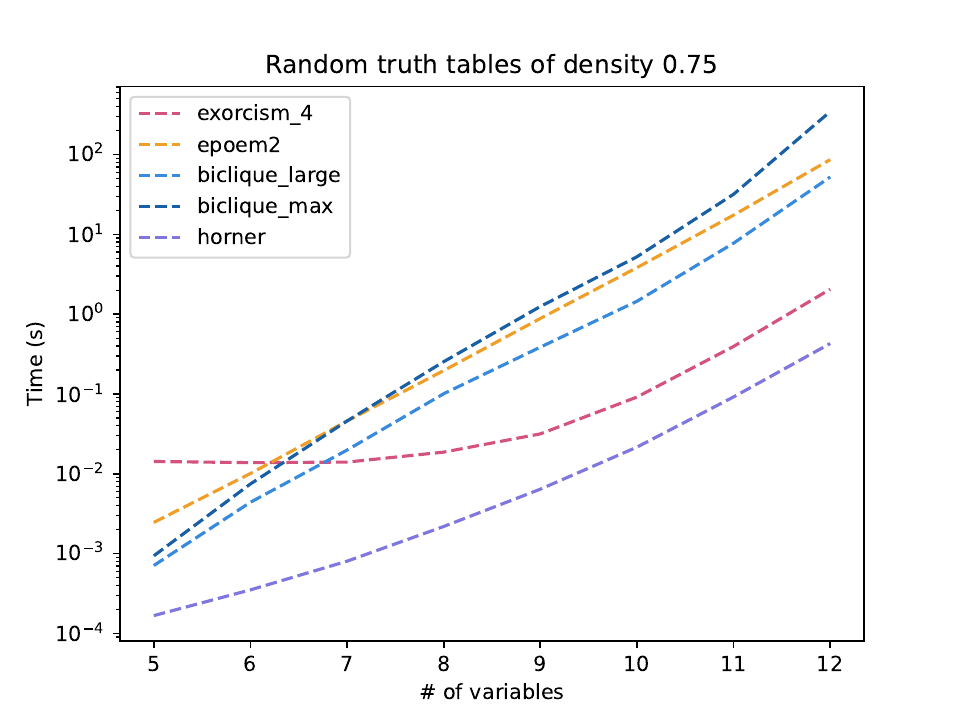}
        \caption{}
    \end{subfigure}
    \caption{Running time of the different heuristics on random truth tables. Each point is averaged over 10 random truth tables. Here we exclude the polynomial encoding (which is instantaneous for those sizes}
    \label{fig:time_random}
\end{figure}

We evaluated all algorithms on truth tables over $n \in [5, 12]$ variables
with $p \cdot 2^n$ ones, for $p \in \{0.25, 0.5, 0.75\}$. For each parameter
setting, we tracked \textsc{and} counts and execution time, averaging results
over 10 random truth tables. The results are reported in
Figures~\ref{fig:and_count_random} and~\ref{fig:time_random}.

Several observations stand out. First, the polynomial representation of
Eq.~\ref{eqn:polyRep} alone already yields significant \textsc{and} count
savings over both the \textbf{initial} baseline and the \textbf{epoem2}
heuristic. This advantage grows with truth table density, which is explained by
an increase in monomial collisions that drastically reduces the total monomial
count relative to the number of ones.

Second, pairing the polynomial encoding with the naive \textbf{horner}
generalization produces results competitive with the state-of-the-art
\textbf{exorcism\_4} method. Combined with the running-time results of
Figure~\ref{fig:time_random}---where \textbf{horner} is substantially faster
than \textbf{exorcism\_4}---this makes \textbf{horner} a strong drop-in
replacement for most practical applications.

Finally, \textbf{biclique\_max} and \textbf{biclique\_large} achieve the
largest \textsc{and} count reductions across all benchmarked methods, with
roughly a $25{\times}$ improvement over the \textbf{initial} baseline and a
$5{\times}$ improvement over \textbf{exorcism\_4}. The two methods perform
similarly, with a slight edge for \textbf{biclique\_max}. In terms of running
time, however, \textbf{biclique\_large} vastly outperforms \textbf{biclique\_max},
making it the most practical choice overall.

\section{Conclusion and future directions}

In this paper we have developed novel algorithms to factor Boolean expressions, expressed in both ESOP and SOP form. The main goal is to optimize the AND-count. For ESOP expression, this quantity is proportional to the number of AND gates or Toffoli gates required to implement a reversible Boolean function with a classical or quantum circuit, respectively. The first algorithm that we develop, is graphical. It reduces the Boolean factoring problem to a variant of the Minimum Biclique Cover of a bipartite graph. Our second algorithm is algebraic and is derived from multi-variate Horner method. From our implementations we find that both our algorithms outperform existing state-of-the-art methods, in time and/or in percentage reduction in the AND-count. Hence they can serve as practical replacements of these methods in applications that involve factoring of Boolean expressions.

One application we are particularly interested in, is synthesis of efficient quantum circuits. Suppose we want to implement a reversible Boolean function with a quantum circuit. Since Toffolis are costly to implement fault-tolerantly, we desire to optimize the Toffoli-count. As explained earlier, we factor the Boolean expression to reduce the AND-count. We can use ancillas to implement each product term and then combine these terms with CNOT or Toffoli. This straight-forward procedure comes at the cost of a lot of extra ancilla qubits. Existing quantum computers as well as the promised early fault-tolerant ones, do not encourage the use of too many qubits. So, though optimization of non-Clifford gates remain a major goal of efficient quantum circuit synthesis, it is unwise to do so at the cost of too many extra qubits. This trade-off will likely hold even in the distant future. 

Also, many interesting quantum circuits, for example arithmetic circuits, have multiple output qubits, each of which implement a reversible Boolean function. These functions usually share the same input variables. In order to optimize the overall AND-count, it might be useful to factor all the functions simultaneously. This might lead to a quantum circuit implementation, with reduced Toffoli-count.

In summary, automatic synthesis of efficient quantum circuits is a challenging task. In the future we hope to apply our factoring algorithms in order to synthesize quantum circuits that implement reversible Boolean functions, while optimizing the Toffoli-count and/or qubit-count.

\section*{Code and Data availability}

The code and relevant data are available at https://github.ibm.com/ibm-q-research/factorust.

\bibliographystyle{alpha}
\bibliography{factor_ref}

@article{1955_Q,
  title="{A way to simplify truth functions}",
  author={Quine, Willard V},
  journal={The American mathematical monthly},
  volume={62},
  number={9},
  pages={627--631},
  year={1955},
  publisher={Taylor \& Francis}
}

@article{1974_HCO,
  title="{MINI: A heuristic approach for logic minimization}",
  author={Hong, Se June and Cain, Robert G. and Ostapko, Daniel L.},
  journal={IBM Journal of Research and Development},
  volume={18},
  number={5},
  pages={443--458},
  year={1974},
  publisher={IBM}
}

@inproceedings{1977_O,
  title="{Contentment in graph theory: covering graphs with cliques}",
  author={Orlin, James},
  booktitle={Indagationes Mathematicae (Proceedings)},
  volume={80},
  number={5},
  pages={406--424},
  year={1977},
  organization={Elsevier}
}

@inproceedings{1982_B,
  title="{The decomposition and factorization of Boolean expressions}",
  author={BRAYTON, Robert K},
  booktitle={Proc. International Symposium on Circuits and Systems},
  pages={49--54},
  year={1982}
}

@book{1984_BHMS,
  title="{Logic minimization algorithms for VLSI synthesis}",
  author={Brayton, Robert K and Hachtel, Gary D and McMullen, Curt and Sangiovanni-Vincentelli, Alberto},
  volume={2},
  year={1984},
  publisher={Springer Science \& Business Media}
}

@article{1984_S,
  title="{Input variable assignment and output phase optimization of PLA's}",
  author={Sasao},
  journal={IEEE Transactions on Computers},
  volume={100},
  number={10},
  pages={879--894},
  year={1984},
  publisher={IEEE}
}

@inproceedings{1986_Y,
  title="{How to generate and exchange secrets}",
  author={Yao, Andrew Chi-Chih},
  booktitle={27th Annual Symposium on Foundations of Computer Science (SFCS 1986)},
  pages={162--167},
  year={1986},
  organization={IEEE}
}

@article{1987_B,
  title="{Factoring logic functions}",
  author={Brayton, Robert K},
  journal={IBM Journal of Research and Development},
  volume={31},
  number={2},
  pages={187--198},
  year={1987},
  publisher={IBM}
}

@book{1988_K,
  title="{Using if-then-else DAGs for multi-level logic minimization}",
  author={Karplus, Kevin},
  year={1988},
  publisher={Computer Research Laboratory, University of California, Santa Cruz}
}

@article{1990_S,
  title="{On approximate solutions for combinatorial optimization problems}",
  author={Simon, Hans Ulrich},
  journal={SIAM Journal on Discrete Mathematics},
  volume={3},
  number={2},
  pages={294--310},
  year={1990},
  publisher={SIAM}
}

@inproceedings{1992_HS,
  title="{Coalgebraic division for multilevel logic synthesis}",
  author={Hsu, Wen-Jun and Shen, Wen-Zen},
  booktitle={[1992] Proceedings 29th ACM/IEEE Design Automation Conference},
  pages={438--442},
  year={1992},
  organization={IEEE}
}

@article{1993_JR,
  title="{Minimal NFA problems are hard}",
  author={Jiang, Tao and Ravikumar, Bala},
  journal={SIAM Journal on Computing},
  volume={22},
  number={6},
  pages={1117--1141},
  year={1993},
  publisher={SIAM}
}

@article{1993_S,
  title="{An exact minimization of AND-EXOR expressions using BDD's}",
  author={Sasao, Tsutomu},
  journal={IFIP WG 10.5 Reed-Muller'93, Germany},
  year={1993}
}

@article{1993_S2,
  title="{EXMIN2: A simplification algorithm for exclusive-OR-sum-of-products expressions for multiple-valued-input two-valued-output functions}",
  author={Sasao, Tsutomu},
  journal={IEEE Transactions on Computer-Aided Design of Integrated Circuits and Systems},
  volume={12},
  number={5},
  pages={621--632},
  year={1993},
  publisher={IEEE}
}

@article{1994_LY,
  title="{On the hardness of approximating minimization problems}",
  author={Lund, Carsten and Yannakakis, Mihalis},
  journal={Journal of the ACM (JACM)},
  volume={41},
  number={5},
  pages={960--981},
  year={1994},
  publisher={ACM New York, NY, USA}
}

@inproceedings{1996_G,
  title="{A fast quantum mechanical algorithm for database search}",
  author={Grover, Lov K},
  booktitle={Proceedings of the twenty-eighth Annual ACM Symposium on Theory of Computing},
  pages={212--219},
  year={1996}
}

@article{1996_M,
  title="{On edge perfectness and classes of bipartite graphs}",
  author={M{\"u}ller, Haiko},
  journal={Discrete Mathematics},
  volume={149},
  number={1-3},
  pages={159--187},
  year={1996},
  publisher={Elsevier}
}

@book{1996_SF,
  title="{Representations of discrete functions}",
  author={Sasao, Tsutomu and Fujita, Masahiro},
  year={1996},
  publisher={Springer}
}

@article{1997_K,
  title="{Quantum computations: algorithms and error correction}",
  author={Kitaev, A Yu},
  journal={Russian Mathematical Surveys},
  volume={52},
  number={6},
  pages={1191--1249},
  year={1997}
}

@article{1999_CC,
  title="{Efficient Boolean division and substitution using redundancy addition and removing}",
  author={Chang, Shih-Chieh and Cheng, David Ihsin},
  journal={IEEE Transactions on Computer-Aided Design of Integrated Circuits and Systems},
  volume={18},
  number={8},
  pages={1096--1106},
  year={1999},
  publisher={IEEE}
}

@inproceedings{2000_YCS,
  title="{BDS: A BDD-based logic optimization system}",
  author={Yang, Congguang and Ciesielski, Maciej and Singhal, Vigyan},
  booktitle={Proceedings of the 37th Annual Design Automation Conference},
  pages={92--97},
  year={2000}
}

@article{2001_MP,
  title="{Fast heuristic minimization of exclusive-sums-of-products}",
  author={Mishchenko, Alan and Perkowski, Marek},
  year={2001}
}

@inproceedings{2001_MSP,
  title="{An algorithm for bi-decomposition of logic functions}",
  author={Mishchenko, Alan and Steinbach, Bernd and Perkowski, Marek},
  booktitle={Proceedings of the 38th Annual Design Automation Conference},
  pages={103--108},
  year={2001}
}

@book{2002_GJ,
  title="{Computers and intractability}",
  author={Garey, Michael R and Johnson, David S},
  volume={29},
  year={2002},
  publisher={Wh Freeman New York}
}

@article{2002_KHP,
  title="{A minimal universal test set for self-test of EXOR-sum-of-products circuits}",
  author={Kalay, Ugur and Hall, Douglas V and Perkowski, Marek A},
  journal={IEEE Transactions on Computers},
  volume={49},
  number={3},
  pages={267--276},
  year={2002},
  publisher={IEEE}
}

@article{2002_SS,
  title="{Boolean division and factorization using binary decision diagrams}",
  author={Stanion, Ted and Sechen, Carl},
  journal={IEEE Transactions on Computer-Aided Design of Integrated Circuits and Systems},
  volume={13},
  number={9},
  pages={1179--1184},
  year={2002},
  publisher={IEEE}
}

@book{2003_GG,
  title="{Modern computer algebra}",
  author={Von Zur Gathen, Joachim and Gerhard, J{\"u}rgen},
  year={2003},
  publisher={Cambridge University Press}
}

@article{2003_P,
  title="{The maximum edge biclique problem is NP-complete}",
  author={Peeters, Ren{\'e}},
  journal={Discrete Applied Mathematics},
  volume={131},
  number={3},
  pages={651--654},
  year={2003},
  publisher={Elsevier}
}

@article{2004_BRSW,
  title="{MIS: A multiple-level logic optimization system}",
  author={Brayton, Robert K and Rudell, Richard and Sangiovanni-Vincentelli, Alberto and Wang, Albert R},
  journal={IEEE Transactions on Computer-Aided Design of Integrated Circuits and Systems},
  volume={6},
  number={6},
  pages={1062--1081},
  year={2004},
  publisher={IEEE}
}

@article{2005_BK,
  title="{Universal quantum computation with ideal Clifford gates and noisy ancillas}",
  author={Bravyi, Sergey and Kitaev, Alexei},
  journal={Physical Review A - Atomic, Molecular, and Optical Physics},
  volume={71},
  number={2},
  pages={022316},
  year={2005},
  publisher={APS}
}

@article{2005_DN,
  title="{The Solovay-Kitaev algorithm}",
  author={Dawson, Christopher M and Nielsen, Michael A},
  journal={arXiv preprint quant-ph/0505030},
  year={2005}
}

@article{2005_MG,
  title="{Factoring Boolean functions using graph partitioning}",
  author={Mintz, Aviad and Golumbic, Martin Charles},
  journal={Discrete Applied Mathematics},
  volume={149},
  number={1-3},
  pages={131--153},
  year={2005},
  publisher={Elsevier}
}

@article{2006_GMR,
  title="{Factoring and recognition of read-once functions using cographs and normality and the readability of functions associated with partial k-trees}",
  author={Golumbic, Martin Charles and Mintz, Aviad and Rotics, Udi},
  journal={Discrete Applied Mathematics},
  volume={154},
  number={10},
  pages={1465--1477},
  year={2006},
  publisher={Elsevier}
}

@article{2006_LSY,
  title="{Finding biclusters by random projections}",
  author={Lonardi, Stefano and Szpankowski, Wojciech and Yang, Qiaofeng},
  journal={Theoretical Computer Science},
  volume={368},
  number={3},
  pages={217--230},
  year={2006},
  publisher={Elsevier}
}

@inproceedings{2006_SLY,
  title="{A note on broadcast encryption key management with applications to large scale emergency alert systems}",
  author={Shu, Guoqiang and Lee, David and Yannakakis, Mihalis},
  booktitle={Proceedings 20th IEEE International Parallel \& Distributed Processing Symposium},
  pages={8--pp},
  year={2006},
  organization={IEEE}
}

@inproceedings{2007_GH,
  title="{Inapproximability of nondeterministic state and transition complexity assuming P$\ne$ NP}",
  author={Gruber, Hermann and Holzer, Markus},
  booktitle={International Conference on Developments in Language Theory},
  pages={205--216},
  year={2007},
  organization={Springer}
}

@article{2007_KPVetal,
  title="{Rectangle Covering Factorization of ESOPs Into Scan-Based Levelized Circuits with Universal Test Set}",
  author={Kalay, Ugur and Perkowski, Marek A and Hall, Douglas V and Steinbach, B and Shahjahan, Shah Amran},
  year={2007}
}

@article{2008_BP,
  title="{Tight bounds for the multiplicative complexity of symmetric functions}",
  author={Boyar, Joan and Peralta, Ren{\'e}},
  journal={Theoretical Computer Science},
  volume={396},
  number={1-3},
  pages={223--246},
  year={2008},
  publisher={Elsevier}
}

@inproceedings{2008_EHMetal,
  title="{Fast exact and heuristic methods for role minimization problems}",
  author={Ene, Alina and Horne, William and Milosavljevic, Nikola and Rao, Prasad and Schreiber, Robert and Tarjan, Robert E},
  booktitle={Proceedings of the 13th ACM Symposium on Access Control Models and Technologies},
  pages={1--10},
  year={2008}
}

@inproceedings{2008_KS,
  title="{Improved garbled circuit: Free XOR gates and applications}",
  author={Kolesnikov, Vladimir and Schneider, Thomas},
  booktitle={International Colloquium on Automata, Languages, and Programming},
  pages={486--498},
  year={2008},
  organization={Springer}
}

@article{2011_CHM,
  title="{Solving circuit optimisation problems in cryptography and cryptanalysis}",
  author={Courtois, Nicolas T and Hulme, Daniel and Mourouzis, Theodosis},
  journal={Cryptology ePrint Archive},
  year={2011}
}

@inproceedings{2011_DFW,
  title="{Improving ESOP-based synthesis of reversible logic using evolutionary algorithms}",
  author={Drechsler, Rolf and Finder, Alexander and Wille, Robert},
  booktitle={European Conference on the Applications of Evolutionary Computation},
  pages={151--161},
  year={2011},
  organization={Springer}
}

@article{2013_BMP,
  title="{Logic minimization techniques with applications to cryptology}",
  author={Boyar, Joan and Matthews, Philip and Peralta, Ren{\'e}},
  journal={Journal of Cryptology},
  volume={26},
  number={2},
  pages={280--312},
  year={2013},
  publisher={Springer}
}

@inproceedings{2014_CHHK,
  title="{Nearly tight approximability results for minimum biclique cover and partition}",
  author={Chalermsook, Parinya and Heydrich, Sandy and Holm, Eugenia and Karrenbauer, Andreas},
  booktitle={European Symposium on Algorithms},
  pages={235--246},
  year={2014},
  organization={Springer}
}

@inproceedings{2014_F,
  title="{On the complexity of computing two nonlinearity measures}",
  author={Find, Magnus Gausdal},
  booktitle={International Computer Science Symposium in Russia},
  pages={167--175},
  year={2014},
  organization={Springer}
}

@inproceedings{2014_TSP,
  title="{The multiplicative complexity of Boolean functions on four and five variables}",
  author={Turan S{\"o}nmez, Meltem and Peralta, Ren{\'e}},
  booktitle={International Workshop on Lightweight Cryptography for Security and Privacy},
  pages={21--33},
  year={2014},
  organization={Springer}
}

@inproceedings{2015_ARSetal,
  title="{Ciphers for MPC and FHE}",
  author={Albrecht, Martin R and Rechberger, Christian and Schneider, Thomas and Tiessen, Tyge and Zohner, Michael},
  booktitle={Annual International Conference on the Theory and Applications of Cryptographic Techniques},
  pages={430--454},
  year={2015},
  organization={Springer}
}

@inproceedings{2015_SHSetal,
  title="{Tinygarble: Highly compressed and scalable sequential garbled circuits}",
  author={Songhori, Ebrahim M and Hussain, Siam U and Sadeghi, Ahmad-Reza and Schneider, Thomas and Koushanfar, Farinaz},
  booktitle={2015 IEEE Symposium on Security and Privacy},
  pages={411--428},
  year={2015},
  organization={IEEE}
}

@inproceedings{2016_GMO,
  title="{$\{$ZKBoo$\}$: faster $\{$Zero-Knowledge$\}$ for Boolean circuits}",
  author={Giacomelli, Irene and Madsen, Jesper and Orlandi, Claudio},
  booktitle={25th Usenix Security Symposium (Usenix Security 16)},
  pages={1069--1083},
  year={2016}
}

@inproceedings{2016_SYL,
  title="{On finding the maximum edge biclique in a bipartite graph: a subspace clustering approach}",
  author={Shaham, Eran and Yu, Honghai and Li, Xiao-Li},
  booktitle={Proceedings of the 2016 SIAM International Conference on Data Mining},
  pages={315--323},
  year={2016},
  organization={SIAM}
}

@INPROCEEDINGS{2016_TGPC,
  author={Tran, Linh and Gronquist, Addison and Perkowski, Marek and Caughman, John},
  booktitle={2016 IEEE 46th International Symposium on Multiple-Valued Logic (ISMVL)}, 
  title={An Improved Factorization Approach to Reversible Circuit Synthesis Based on EXORs of Products of EXORs}, 
  year={2016},
  volume={},
  number={},
  pages={37-43},
  doi={10.1109/ISMVL.2016.56}}

@inproceedings{2017_CDGetal,
  title="{Post-quantum zero-knowledge and signatures from symmetric-key primitives}",
  author={Chase, Melissa and Derler, David and Goldfeder, Steven and Orlandi, Claudio and Ramacher, Sebastian and Rechberger, Christian and Slamanig, Daniel and Zaverucha, Greg},
  booktitle={Proceedings of the 2017 ACM Sigsac Conference on Computer and Communications Security},
  pages={1825--1842},
  year={2017}
}

@inproceedings{2017_PRS,
  title="{REVS: A tool for space-optimized reversible circuit synthesis}",
  author={Parent, Alex and Roetteler, Martin and Svore, Krysta M},
  booktitle={International Conference on Reversible Computation},
  pages={90--101},
  year={2017},
  organization={Springer}
}

@article{2017_SP,
  title="{Compact XOR Bi-Decomposition for Generalized Lattices of Boolean Functions}",
  author={Steinbach, Bernd and Posthoff, Christian},
  journal={Proc. RMW},
  year={2017}
}

@article{2017_SRWM,
  title="{Logic synthesis for quantum computing}",
  author={Soeken, Mathias and Roetteler, Martin and Wiebe, Nathan and De Micheli, Giovanni},
  journal={arXiv preprint arXiv:1706.02721},
  year={2017}
}

@inproceedings{2018_MSRetal,
  title="{A best-fit mapping algorithm to facilitate ESOP-decomposition in Clifford+T quantum network synthesis}",
  author={Meuli, Giulia and Soeken, Mathias and Roetteler, Martin and Wiebe, Nathan and De Micheli, Giovanni},
  booktitle={2018 23rd Asia and South Pacific Design Automation Conference (ASP-DAC)},
  pages={664--669},
  year={2018},
  organization={IEEE}
}

@inproceedings{2018_SO,
  title="{Finding maximum edge biclique in bipartite networks by integer programming}",
  author={S{\"o}zdinler, Melih and {\"O}zturan, Can},
  booktitle={2018 IEEE International Conference on Computational Science and Engineering (CSE)},
  pages={132--137},
  year={2018},
  organization={IEEE}
}

@article{2019_BDLK,
  title="{Optimal usage of quantum random access memory in quantum machine learning}",
  author={Bang, Jeongho and Dutta, Arijit and Lee, Seung-Woo and Kim, Jaewan},
  journal={Physical Review A},
  volume={99},
  number={1},
  pages={012326},
  year={2019},
  publisher={APS}
}

@article{2019_CSP,
  title="{The multiplicative complexity of 6-variable Boolean functions}",
  author={{\c{C}}al{\i}k, {\c{C}}a{\u{g}}da{\c{s}} and S{\"o}nmez Turan, Meltem and Peralta, Ren{\'e}},
  journal={Cryptography and Communications},
  volume={11},
  number={1},
  pages={93--107},
  year={2019},
  publisher={Springer}
}

@inproceedings{2019_MSCetal,
  title="{The role of multiplicative complexity in compiling low T-count oracle circuits}",
  author={Meuli, Giulia and Soeken, Mathias and Campbell, Earl and Roetteler, Martin and De Micheli, Giovanni},
  booktitle={2019 IEEE/ACM International Conference on Computer-Aided Design (ICCAD)},
  pages={1--8},
  year={2019},
  organization={IEEE}
}

@inproceedings{2019_MSEetal,
  title="{Evaluating ESOP optimization methods in quantum compilation flows}",
  author={Meuli, Giulia and Schmitt, Bruno and Ehlers, R{\"u}diger and Riener, Heinz and De Micheli, Giovanni},
  booktitle={Reversible Computation: 11th International Conference, RC 2019, Lausanne, Switzerland, June 24--25, 2019, Proceedings 11},
  pages={191--206},
  year={2019},
  organization={Springer}
}

@inproceedings{2019_TSAM,
  title="{Reducing the multiplicative complexity in logic networks for cryptography and security applications}",
  author={Testa, Eleonora and Soeken, Mathias and Amar{\`u}, Luca and De Micheli, Giovanni},
  booktitle={Proceedings of the 56th Annual Design Automation Conference 2019},
  pages={1--6},
  year={2019}
}

@article{2020_AGGW,
  title="{Convex optimization using quantum oracles}",
  author={van Apeldoorn, Joran and Gily{\'e}n, Andr{\'a}s and Gribling, Sander and de Wolf, Ronald},
  journal={Quantum},
  volume={4},
  pages={220},
  year={2020},
  publisher={Verein zur F{\"o}rderung des Open Access Publizierens in den Quantenwissenschaften}
}

@article{2020_BCHK,
  title="{Lower bounds on the non-Clifford resources for quantum computations}",
  author={Beverland, Michael and Campbell, Earl and Howard, Mark and Kliuchnikov, Vadym},
  journal={Quantum Science \& Technology},
  volume={5},
  number={3},
  pages={035009},
  year={2020},
  publisher={IOP Publishing}
}

@article{2020_HS,
  title="{Lowering the T-depth of quantum circuits by reducing the multiplicative depth of logic networks}",
  author={H{\"a}ner, Thomas and Soeken, Mathias},
  journal={arXiv preprint arXiv:2006.03845},
  year={2020}
}

@article{2020_LQLetal,
  title="{Maximum biclique search at billion scale}",
  author={Lyu, Bingqing and Qin, Lu and Lin, Xuemin and Zhang, Ying and Qian, Zhengping and Zhou, Jingren},
  journal={Proceedings of the VLDB Endowment},
  year={2020},
  publisher={ASSOC COMPUTING MACHINERY}
}

@article{2020_S,
  title={Determining the multiplicative complexity of boolean functions using SAT},
  author={Soeken, Mathias},
  journal={arXiv preprint arXiv:2005.01778},
  year={2020}
}

@inproceedings{2020_TSRetal,
  title={A logic synthesis toolbox for reducing the multiplicative complexity in logic networks},
  author={Testa, Eleonora and Soeken, Mathias and Riener, Heinz and Amar{\`u}, Luca Gaetano and De Micheli, Giovanni},
  booktitle={Proceedings Of The 2020 Design, Automation \& Test In Europe Conference \& Exhibition (Date 2020)},
  pages={568--573},
  year={2020},
  organization={IEEE}
}

@article{2021_BLHetal,
  title="{Quantum computing enhanced computational catalysis}",
  author={von Burg, Vera and Low, Guang Hao and H{\"a}ner, Thomas and Steiger, Damian S and Reiher, Markus and Roetteler, Martin and Troyer, Matthias},
  journal={Physical Review Research},
  volume={3},
  number={3},
  pages={033055},
  year={2021},
  publisher={APS}
}

@inproceedings{2021_OCKK,
  title="{Quantum convolutional neural network for resource-efficient image classification: A quantum random access memory (QRAM) approach}",
  author={Oh, Seunghyeok and Choi, Jaeho and Kim, Jong-Kook and Kim, Joongheon},
  booktitle={2021 International Conference on Information Networking (ICOIN)},
  pages={50--52},
  year={2021},
  organization={IEEE}
}

@article{2023_MWZ,
  title="{Synthesizing efficient circuits for Hamiltonian simulation}",
  author={Mukhopadhyay, Priyanka and Wiebe, Nathan and Zhang, Hong Tao},
  journal={npj Quantum Information},
  volume={9},
  number={1},
  pages={31},
  year={2023},
  publisher={Nature Publishing Group UK London}
}

@article{2023_SBBetal,
  title="{Automatic generation of Grover quantum oracles for arbitrary data structures}",
  author={Seidel, Raphael and Becker, Colin Kai-Uwe and Bock, Sebastian and Tcholtchev, Nikolay and Gheorghe-Pop, Ilie-Daniel and Hauswirth, Manfred},
  journal={Quantum Science and Technology},
  volume={8},
  number={2},
  pages={025003},
  year={2023},
  publisher={IOP Publishing}
}

@article{2024_Mtof,
  title="{Synthesizing Toffoli-optimal quantum circuits for arbitrary multi-qubit unitaries}",
  author={Mukhopadhyay, Priyanka},
  journal={arXiv preprint arXiv:2401.08950},
  year={2024}
}

@article{2024_YM,
  title="{Expediting Homomorphic Computation via Multiplicative Complexity-aware Multiplicative Depth Minimization}",
  author={Yu, Mingfei and De Micheli, Giovanni},
  journal={Cryptology ePrint Archive},
  year={2024}
}

@article{2025_CLDetal,
  title="{On the Efficient Discovery of Maximum k-Defective Biclique}",
  author={Cui, Donghang and Li, Ronghua and Dai, Qiangqiang and Qin, Hongchao and Wang, Guoren},
  journal={arXiv preprint arXiv:2506.16121},
  year={2025}
}

@article{2025_GB,
  title="{Improved binary linear programming models for finding maximum edge Bi-clique in bipartite graphs}",
  author={Ghadiri, Mohammad Javad and Bagherian, Mehri},
  journal={The Journal of Supercomputing},
  volume={81},
  number={1},
  pages={356},
  year={2025},
  publisher={Springer}
}

@article{2025_Mqram,
  title="{A quantum random access memory (QRAM) using a polynomial encoding of binary strings}",
  author={Mukhopadhyay, Priyanka},
  journal={Scientific Reports},
  volume={15},
  number={1},
  pages={11002},
  year={2025},
  publisher={Nature Publishing Group UK London}
}

@article{2026_HMAetal,
  title="{Scattering Processes from Quantum Simulation Algorithms for Scalar Field Theories}",
  author={Hardy, Andrew and Mukhopadhyay, Priyanka and Alam, M Sohaib and Konik, Robert and Hormozi, Layla and Rieffel, Eleanor and Hadfield, Stuart and Barata, Jo{\~a}o and Venugopalan, Raju and Kharzeev, Dmitri E and others},
  journal={PRX Quantum},
  volume={7},
  number={1},
  pages={010343},
  year={2026},
  publisher={APS}
}

@article{pena2000multivariate,
    author  = {Pe{\~n}a, J. M. and Sauer, T.},
    title   = {On the Multivariate {H}orner Scheme},
    journal = {SIAM Journal on Numerical Analysis},
    volume  = {37},
    number  = {4},
    pages   = {1186--1197},
    year    = {2000},
    doi     = {10.1137/S0036142998347521}
  }


\end{document}